\documentclass[manuscript,screen]{acmart}

\AtBeginDocument{%
 \providecommand\BibTeX{{%
 \normalfont B\kern-0.5em{\scshape i\kern-0.25em b}\kern-0.8em\TeX}}}

\usepackage{todonotes}
\usepackage{placeins}
\usepackage{subcaption}

\begin{document}

\title[Robust Markov stability for community detection at a scale learned based on the structure]{Robust Markov stability for community detection at a scale learned based on the structure}


\author{Samin Aref}
\authornote{Corresponding author's email: aref@mie.utoronto.ca}
\orcid{0000-0002-5870-9253}
\affiliation{%
  \institution{Department of Mechanical and Industrial Engineering, University of Toronto}
  \streetaddress{5 King’s College Road}
  \city{Toronto}
  \state{ON}
  \country{Canada}
  \postcode{M5S 3G8}
}

\author{Sanchaai Mathiyarasan}
\orcid{0009-0000-6784-9445}
\affiliation{%
  \institution{Department of Computer Science, University of Toronto}
  \streetaddress{40 St. George Street}
  \city{Toronto}
  \state{ON}
  \country{Canada}
  \postcode{M5S 2E4}
}

\renewcommand{\shortauthors}{Aref and Mathiyarasan}

\begin{abstract}
Community detection, the unsupervised task of clustering nodes of a graph, finds applications across various fields. The common approaches for community detection involve optimizing an objective function to partition the nodes into communities at a single scale of granularity. However, the single-scale approaches often fall short of producing partitions that are robust and at a suitable scale. The existing algorithm, PyGenStability, returns multiple robust partitions for a network by optimizing the multi-scale Markov stability function. However, in cases where the suitable scale is not known or assumed by the user, there is no principled method to select a single robust partition at a suitable scale from the multiple partitions that PyGenStability produces. 

Our proposed method combines the Markov stability framework with a pre-trained machine learning model for scale selection to obtain one robust partition at a scale that is learned based on the graph structure. This automatic scale selection involves using a gradient boosting model pre-trained on hand-crafted and embedding-based network features from a labeled dataset of 10k benchmark networks. This model was trained to predicts the scale value that maximizes the similarity of the output partition to the planted partition of the benchmark network. Combining our scale selection algorithm with the PyGenStability algorithm results in PyGenStabilityOne (PO): a hyperparameter-free multi-scale community detection algorithm that returns one robust partition at a suitable scale without the need for any assumptions, input, or tweaking from the user. We compare the performance of PO against 29 algorithms and show that it outperforms 25 other algorithms by statistically meaningful margins. Our results facilitate choosing between community detection algorithms, among which PO stands out as the accurate, robust, and hyperparameter-free method.
\end{abstract}



\received{\today}

\maketitle

\section{Introduction}
\label{s:intro}
Community detection (CD), the unsupervised task of partitioning the nodes of a graph \citep{schaub2017many}, has been a fundamental challenge in various academic fields, including social network analysis and social computing \citep{fortunato2022newman}. Common optimization approaches for CD aim to optimize a network-level objective function at a specific scale of granularity (resolution), such as single-resolution modularity \citep{newman_modularity_2006,fortunato2016,aref2023suboptimality}, surprise \citep{surprise_2015,marchese2022detecting}, modularity density \citep{sato_enhanced_2019}, and single-scale Markov stability \citep{delvenne2010_time_scale_stability, Lambiotte2014_markov_processes}. Despite their widespread adoption in numerous peer-reviewed studies \citep{Kosowski2020,fortunato2022newman}, these single-resolution algorithms carry a notable risk of failing to identify communities that are robust and at a suitable scale/resolution \citep{kawamoto2019counting}. They may produce partitions that are substantially different from the dominant underlying community structure \citep{good_performance_2010,dinh_network_2015,newman_equivalence_2016,aref2023analyzing} partly due to optimizing a function at a scale that is not consistent with the structure of the input graph.

There are various limitations associated with modularity \citep{peixoto_2023} and other single-scale objective functions \citep{rosvall_2007,surprise_2015}, including the presence of a resolution limit \citep{fortunato_2007}. These limitations have motivated researchers to explore alternative methods \citep{sato_enhanced_2019,bpp2020,aref2022bayan,marchese2022detecting,paul_community_2022,liu_community_2023}, including multi-scale\footnote{A multi-scale method involves searching through multiple scales of granularity (multiple resolutions) for obtaining partitions.} Markov stability \citep{Schaub2019}, for identifying communities by exploring candidate partitions at multiple scales (multiple resolutions). Optimizing multi-scale Markov stability was shown to resolve several limitations of modularity \citep{Schaub2019}. However, Markov stability has a scale parameter, \textit{Markov time}, whose selection poses the same challenges as determining the resolution parameter in a multi-resolution modularity maximization scheme \citep{lancichinetti_limits_2011}. 

The landscape of CD algorithms is vast, encompassing hundreds of approaches \citep{fortunato2022newman}, with dozens readily available in public libraries such as \textit{NetworkX} \citep{networkx}, \textit{graph-tool} \citep{graph-tool_2014}, \textit{CDLib} \citep{rossetti2019cdlib}, and the \textit{PyGenStability} library \citep{pygenstability_2023}. New CD algorithms are often compared to just a few other methods on benchmark datasets. The lack of comprehensive assessments leads to the use of convenient or widely adopted \citep{Kosowski2020} algorithms \citep{cscw14_community,cscw21_community} that may not ideally be suited for the specific CD task at hand without imposing additional assumptions or user input requirements. Using a traditional single-scale method will return a partition at a scale that is hard-coded in the method\footnote{A single-scale modularity maximization always returns a partition at the modularity resolution $\gamma=1$ regardless of the input network.} rather than the scale that is aligned with the structure of the input network. We propose a multi-scale CD algorithm that returns a robust partition at a scale that is learned based on the structure of the input network without any assumption required from the user or any hyperparameters to be tuned/adjusted by the user. This involves building upon the PyGenStability algorithm to solve a problem that it currently does not support. We compare our extended algorithm (PyGenStabilityOne - PO) with 29 other algorithms on 500 synthetic benchmark graphs and five real networks. This approach allows us to achieve two goals: (1) identify the advantages and disadvantages of existing CD algorithms, irrespective of whether they employ Markov stability, and (2) assess the performance and applicability of PO against a comprehensive set of performance baselines.

This paper is structured as follows: We briefly describe our proposed algorithm in Section \ref{s:proposed}. We review existing algorithms in Section \ref{s:review}. The technical background and underpinnings of our algorithm are explained in Section \ref{s:technical}. The machine learning core of our proposed method is described in Section \ref{s:ml}. The results from comparing our proposed algorithm to existing baselines are provided in Section \ref{s:result}. Section \ref{ss:cscw} provides some examples on using community detection in CSCW research. Finally, Section \ref{s:discuss} provides a discussion on our main results and concludes the study.

\subsection{Preliminaries and notations}

We denote the simple undirected and unweighted\footnote{We restrict this study to undirected and unweighted graphs because they are considered legal input for all the 30 CD algorithms that we would like to compare.} input graph as $G=(V,E)$, where $V$ is the set of nodes and $E$ is the set of edges. Graph $G$ consists of $|V|=n$ nodes (vertices) and $|E|=m$ edges. The symmetric adjacency matrix of graph $G$ is denoted by $\textbf{A}=[a_{ij}]$, with the entry at row $i$ and column $j$ being $a_{ij}$. The entry $a_{ij}$ signifies whether node $i$ is connected to node $j$ ($a_{ij}=1$) or not ($a_{ij}=0$). The degree of node $i$ is represented as $d_i=\sum_j{a_{ij}}$. The node set $V$ of the input graph $G$ can be \textit{partitioned into} (i.e., divided into an unspecified\footnote{This paper addresses the more general CD problem where the number of communities, $c$, is not specified by the user.} number of non-overlapping) communities based on the partition $P=\{V_1,V_2, \dots, V_c \}$.

\section{The proposed algorithm}
\label{s:proposed}
We propose the PyGenStabilityOne (PO) algorithm which returns a single \textbf{robust} partition at a \textbf{suitable scale} that is consistent with the structure of the network. The key practical advantage of our proposed method is that no information/assumption about the partition (number of communities, scale, resolution) is required from the user. For the robustness of the partition, we rely on the recently developed algorithm, PyGenStability \citep{pygenstability_2023} and its robust scale selection heuristic \citep{schindler2023multiscale}. The PyGenStability algorithm does not automatically return a single partition at a suitable scale that is consistent with the structure of the input network. Selecting one partition from the multiple partitions returned by PyGenStability is not a straightforward task. While the Markov stability and PyGenStability are both based on obtaining multiple partitions for an input network, our contribution is in developing an extension for the PyGenStability algorithm (called PyGenStabilityOne) that automatically returns one partition at a suitable scale after solving the multi-scale community detection problem. 

\begin{figure}[htpb!]
 \includegraphics[width=\textwidth]{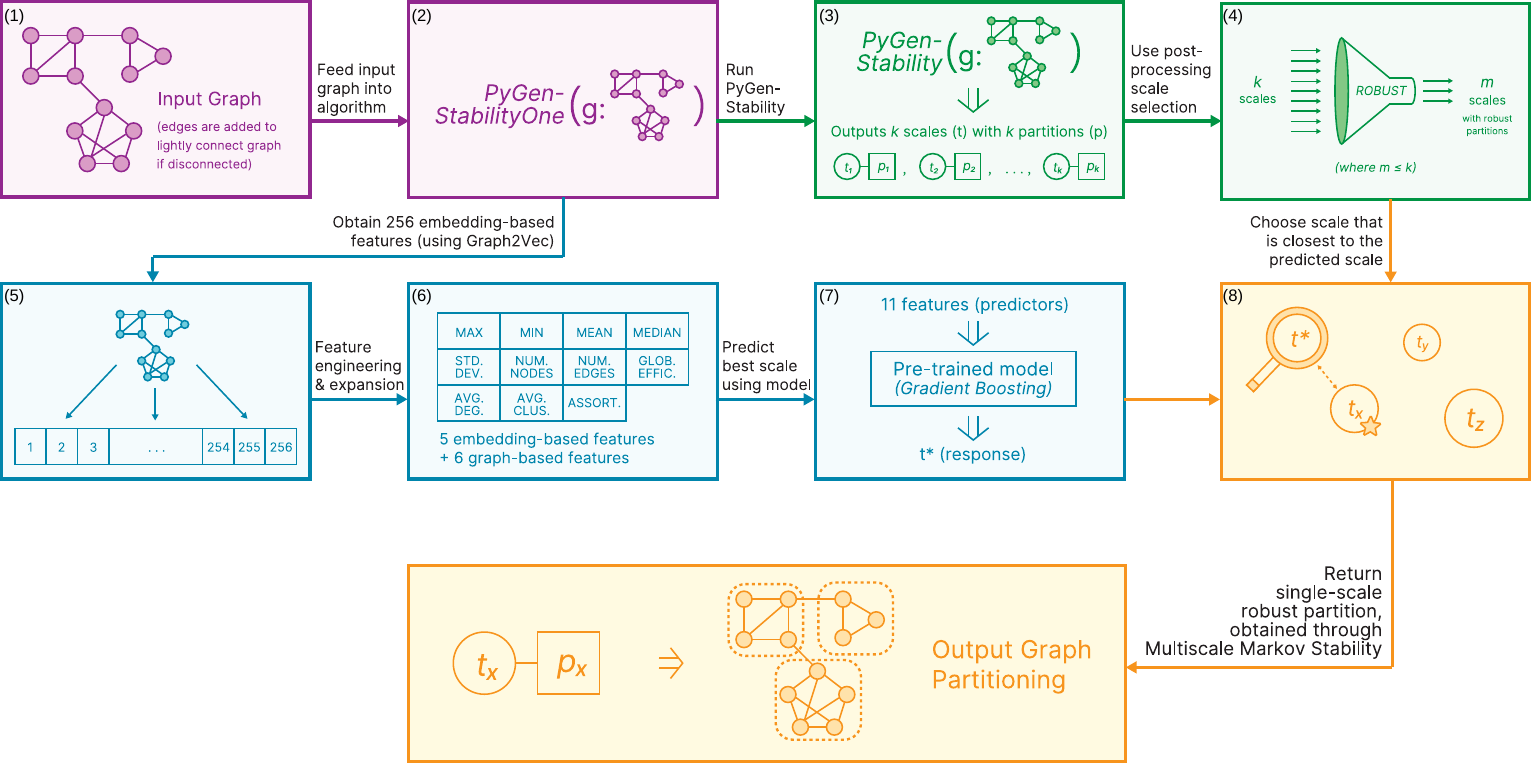}
 \caption{Flowchart of the eight steps within the PyGenStabilityOne (PO) algorithm to take an input graph and return a single robust partition at a suitable scale that is consistent with the structure of the graph. (Magnify the high-resolution figure on screen for details.)}
 \label{fig:flowchart}
        \Description[<short description>]{<long description>}  
\end{figure}

Fig.~\ref{fig:flowchart} illustrates the eight building blocks of our proposed algorithm, which are the steps it takes to convert the input (the graph) into the output (the robust partition). In what follows the proposed algorithm is described in simple terms. 

\noindent
\textbf{Step (1) - Pre-processing:}
For any input graph that is disconnected, a pre-processing step (described in Section \ref{ss:preprocess}) lightly connects the multiple components of the graph by adding the minimum required edges.\\
\textbf{Step (2) - Function call:}
The PyGenStabilityOne algorithm is called for the pre-processed input graph. As a result, two independent processes are initiated in parallel.\\
\textbf{Step (3) - PyGenStability:}
The PyGenStability algorithm (described in Section \ref{ss:pygen}) is used to obtain a series of $k$ pairs of scale values $t_i$ and partitions $p_i$.\\
\textbf{Step (4) - Robust partition heuristic:}
The post-processing heuristic of PyGenStability (described in Section \ref{ss:heuristic}) is used to screen the $k$ partitions into a fewer number $m$ of partitions that satisfy the definition of robustness.\\
\textbf{Step (5) - Embedding:}
A 256-wide array of embedding is extracted (as described in Section \ref{ss:feature}) from the input graph using the Graph2Vec algorithm \citep{narayanangraph2vec}.\\
\textbf{Step (6) - Feature engineering:}
The 256 features are engineered using five summary statistics and are expanded (as described in Section \ref{ss:feature}) using six graph measurements to create 11 features for supervised learning.\\
\textbf{Step (7) - Prediction:}
Using a pre-trained machine learning model (described in Section \ref{s:ml}), the 11 features are used to predict the scale value $t^*$ whose corresponding partition is aligned with the structure of the graph. \\
\textbf{Step (8) - Output:}
The robust partition $p_x$ whose corresponding scale $t_x$ is the closest to the predicted scale $t^*$ is returned as the final output of the PyGenStabilityOne algorithm.

Additional background and the technical underpinnings for the proposed algorithm are described in Section \ref{s:technical}. In Section \ref{s:review}, we review the existing CD algorithms that are most relevant to the definition of our unsupervised problem: the non-overlapping clustering (i.e., partitioning) of nodes purely based on the structure of the input network where the number of clusters (communities) is not known or assumed.

\section{Reviewing existing approaches for community detection}
\label{s:review}
We review and assess 29 other algorithms spanning from 1970 to 2023 which cover (to the best of our knowledge) the most comprehensive set of baseline methods for proposing a new CD algorithm. Our review and assessment include eight modularity-based heuristics, seven non-modularity based optimization methods, four inferential methods, three walk-based methods, three propagation-based methods, two graph cutting methods, and two naive adaptations of the Markov stability method.

The 29 existing algorithms can be grouped into seven categories as briefly discussed in the following seven subsections. Readers who are more interested in the proposed method and how it compares to existing methods, may directly skip to Section \ref{s:technical} or Section \ref{s:ml}. For each algorithm, a two-letter indicator is provided in square brackets [ ]. For some algorithm a shortened or abbreviated name is also provided in round brackets ( ).

\subsection{Modularity-based optimization algorithms}\

\textbf{Clauset-Newman-Moore (CNM) [CN]:}
The Clauset-Newman-Moore greedy algorithm starts with each node being its own community. It subsequently employs a greedy approach to merge two communities such that the merging has the largest immediate positive contribution to modularity \citep{clauset_finding_2004}.

\textbf{Louvain [LO]:}
The Louvain algorithm consists of two sets of iterative steps: (1) moving nodes to locally increase modularity and (2) aggregating communities identified in the first step \citep{blondel_fast_2008}. Despite being the most widely used CD algorithm \citep{Kosowski2020}, it may occasionally result in disconnected components within the same community \citep{traag_louvain_2019}.

\textbf{Leiden [LE]:}
The Leiden algorithm is designed based off of the Louvain algorithm. Leiden aims to address a limitation of the Louvain algorithm by ensuring well-connected communities, where all subsets of communities are locally optimally assigned \citep{traag_louvain_2019}. The Leiden algorithm can be used for objective functions other than modularity. In this article, however, Leiden specifically refers to the Leiden algorithm for modularity maximization.

\textbf{Reichardt-Bornholdt with configuration model [RB]:}
This algorithm is based on the modularity function discussed by Reichardt and Bornholdt which has the configuration model as the null model (the standard null model for the modularity function). It employs the same heuristic principles as the Leiden algorithm, but it also handles weighted and directed graphs \citep{rb_pots_2008}.

\textbf{Combo [CO]:}
The Combo algorithm is a versatile method that supports several graph optimization problems including heuristic modularity maximization. It includes two iterative processes: (1) identifying optimal mergers, splits, or recombinations of communities to increase modularity and (2) applying a sequence of Kernighan-Lin bisections \citep{kernighan1970efficient} to communities such that modularity is further increased \citep{sobolevsky2014general}.

\textbf{Belief [BE]:}
The Belief propagation algorithm aims to achieve consensus among different high-modularity partitions using a message-passing approach \citep{zhang2014}. It relies on the assumption that modularity maximization attempts often result in numerous competing partitions that are poorly correlated.

\textbf{Paris [PA]:}
The Paris algorithm has a sliding resolution for modularity and aims to capture the multi-scale community structure of real networks without requiring a resolution parameter. It generates a hierarchical community structure by assessing the distance between communities using a nearest-neighbor chain approach \citep{paris_2018}.

\textbf{EdMot [ED]:}
The EdMot-Louvain algorithm (EdMot for short) is designed to address the hypergraph fragmentation problem found in earlier motif-based community detection methods \citep{edmot_2019}. It initially constructs a graph representation of higher-order motifs (small, dense subgraph patterns) and subsequently partitions this graph using the Louvain method. The goal is to heuristically maximize modularity based on these higher-order motifs \citep{edmot_2019}.

\subsection{Non-modularity based optimization algorithms}\

\textbf{Reichardt-Bornholdt with Erd\H{o}s-R\'{e}nyi [RE]:}
This algorithm is based on the modularity function discussed by Reichardt and Bornholdt which has the Erd\H{o}s-R\'{e}nyi model as the null model \citep{rber_pots_2006}. It employs the same heuristic principles as the Leiden algorithm, but it aims to optimize a modified version of the modularity function whose null model is an Erd\H{o}s-R\'{e}nyi model instead of the configuration model.

\textbf{Genetic algorithm [GA]:}
The GA-net algorithm (GA for short) \citep{ga_2008} uses the meta-heuristic method, genetic algorithm, for detecting communities through optimizing an objective function called \textit{community score}.

\textbf{Constant Potts model (CPM) [CP]:}
The CPM algorithm \citep{cpm_2011} heuristically optimizes an objective function based on the first principle Potts model \citep{Reichardt_2007}. It is suggested to be a resolution-limit-free method which is desirable at least from a theoretical standpoint.

\textbf{Significant scales [SC]:}
The significant scales (a.k.a significant communities) algorithm \citep{significance_communities_2013} relies on the principle of scanning a range of resolutions for a new notion of significance: the probability that a partition with a required minimum number of internal edges can be found in a random graph. It is initially built on top of the CPM method. In its more general form, however, this algorithm is independent of the objective function and is directly related to the notion of \textit{surprise}.

\textbf{Surprise [SU]:}
The surprise algorithm \citep{surprise_2015} is an approximate optimization method for the surprise objective function. While the surprise objective function is motivated by the resolution limit issue of modularity, it suffers from a resolution limit issue of its own. Compared to modularity, optimizing surprise typically leads to detecting more communities.

\textbf{Weighted Community Clustering (WCC) [WC]:}
The WCC algorithm \citep{scd_2014} is designed as a scalable optimization method for a triangle-based objective function known as weighted community clustering. It relies on the assumption that all real networks have a high clustering coefficient (high fraction of closed triangles). It works in three steps: (1) removing edges that are not part of any triangles, (2) generating an initial partition by heuristically optimizing the WCC function, and (3) refining the partition.

\textbf{GemSec [GE]:}
The GemSec algorithm \citep{gemsec_2019} is a node embedding method that also returns clusters for the nodes. It produces an abstract feature space for the nodes such that the node features minimize the negative log-likelihood of preserving sampled node neighborhoods.

\subsection{Inferential algorithms}\

\textbf{Stochastic Block Model (SBM) [SB]:}
The SBM algorithm \citep{sbm_2014} (a.k.a.\ the SBM minimum description length algorithm) is partly motivated by the observation that modularity optimization is only equivalent \citep{newman_equivalence_2016} to a maximum likelihood under very stringent conditions. SBM and other inferential methods, however, are designed with a mindset of making statistical inference about the graph generation process. SBM relies on heuristically minimizing the \textit{description length} which is an information-theoretic measure for the compression of a partition.

\textbf{SBM with Markov Chain  Monte Carlo [SM]:}
SBM with MCMC is a specific way of using the SBM algorithm. It involves running a Markov Chain  Monte Carlo post-processing to improve the results of the SBM minimum description length algorithm.

\textbf{Bayesian Planted Partition (BPP) [BP]:}
The BPP algorithm \citep{bpp2020} is an inferential method that relies on Bayesian statistics. It uses a nonparametric Bayesian formulation of the planted partition to extract an assortative community structure.

\textbf{BPP with Markov Chain  Monte Carlo [BM]:}
BPP with MCMC is a specific way of using the BPP method. It involves running a Markov Chain  Monte Carlo post-processing to improve the results of the BPP algorithm.

\subsection{Walk-based algorithms}\

\textbf{Walktrap [WA]:}
The Walktrap algorithm \citep{Walktrap_2006} uses a walk-based measure of similarity between nodes to capture the communities as dense subgraphs of sparse graphs. It runs in $ \mathcal{O}(mn^2)$ in the worst case, and in $\mathcal{O} (n^2 \log n)$ in most real-world cases.

\textbf{Infomap [IN]:}
The Infomap algorithm \citep{rosvall_2008} is based on an information-theoretic foundation for modularity. It aims to find an optimal compression of the network topology using the structural regularities. It works based on compressing the probability flow of random walks.

\textbf{Diffusion Entropy Reducer (DER) [DE]:}
DER \citep{der_2015} uses random walks to embed the nodes in a space of measures. Then, a modification of the k-means clustering algorithm is applied to the node embedding to obtain clusters (communities). For specific instances, DER runs in $\mathcal{O}(m)$. 

\subsection{Propagation-based algorithms}\

\textbf{Chinese whispers [CW]:}
Chinese whispers \citep{chinesewhispers_2006} is a matrix operation algorithm that runs in $\mathcal{O}(m)$. It is inspired by and named after the children’s game, where each person whispers a message to the next person in the line to ultimately arrive at an altered funny message (a.k.a.\ the telephone game). The algorithm is designed to return clusters of nodes that broadcast the same message to their neighbors.

\textbf{Asynchronous label propagation [AL]:}
Instead of relying on an objective function, this algorithm \citep{asynchronous_label_propagation2007} initializes each node with a unique label. Then, each nodes adopts the most common label from its neighbors. Sequentially, densely connected groups of nodes form a label consensus. The algorithm is suggested to be near linear time, and more precisely it runs in $\Omega(m)$ (i.e.\, the best-case running time is linear in the number of edges).

\textbf{Semi-synchronous label propagation [LP]:}
The semi-synchronous label propagation algorithm \citep{label_propagation_2010} improves earlier synchronous and asynchronous label propagation algorithms. It is motivated by the observations that algorithm termination is not guaranteed in synchronous label propagation; and performance can be the worst in asynchronous label propagation. The semi-synchronous label propagation combines the advantages of the earlier label propagation methods, and is guaranteed to converge to a stable partition through extremely parallelizable propagation steps.

\subsection{Graph cutting algorithms}\

\textbf{Kernighan-Lin bisection [KL]:}
This 1970 graph bisection algorithm \citep{kernighan1970efficient} aims to split the node set into two subsets of nearly equal size, in a way that minimizes the number of cut edges (edges with endpoints in the two subsets). After initialization, the algorithm improves the partition by pairing up nodes from the two subsets in a greedy way, so that swapping the two paired nodes will reduce the number of cut edges. It runs in $\mathcal{O}(n^2 \log n)$.

\textbf{k-cut [KC]:}
k-cut is a spectral algorithm which was designed to return high-modularity partitions through minimizing a normalized cut criterion \citep{kcut_2007}. It works based on computing the second smallest eigenvector of the Laplacian matrix corresponding to the network. It uses a greedy heuristic for recursively partitioning a network into smaller sub-networks as long as modularity increases.

\subsection{Markov stability algorithms}\

\textbf{Markov stability with random partition selection [MR]:}
The MR algorithm is a naïve adaptation of the multi-scale Markov stability algorithm (described in Section \ref{ss:pygen}) from the PyGenStability library \citep{pygenstability_2023}. It involves the post-processing step (described in Section \ref{ss:heuristic}) that uses the robust partition heuristic \citep{schindler2023multiscale}. These two algorithmic steps generate multiple partitions each at a different scale. Then, in MR, a single final partition is randomly selected out of the several candidate partitions.

\textbf{Markov stability with minimum NVI [MV]:}
The MV algorithm also relies on the Markov stability \citep{pygenstability_2023} with the same heuristic post-processing \citep{schindler2023multiscale}, but it is adapted differently to generate a single partition. It differs from MR only in the final partition selection step. In the MV adaptation, the single final partition is chosen based on having the lowest Normalized Variance of Information (NVI) \citep{pygenstability_2023}.

\section{Technical background for the proposed algorithm}
\label{s:technical}

\subsection{Pre-processing step for disconnected input}
\label{ss:preprocess}
As the PyGenStability library requires its input graph to be connected, we include a pre-processing step in PO that creates a lightly connected graph from every disconnected input graph. This procedure iterates over pairs of largest components (ordered by their number of nodes) and connects them by adding an edge between the highest-degree nodes of the two components. In the subsequent iterations, the remaining unmodified nodes that have the highest degrees are used (to ensure each node is modified as little as possible). High degree nodes are chosen because their impact on the Markov stability function is less dramatic than the impact of creating new edges between low-degree nodes. PO uses this pre-processing step for any input that is a disconnected graph.

\subsection{Generalized Markov stability}
The Markov stability \citep{delvenne2010_time_scale_stability} framework relies on graph diffusion processes to obtain \textit{robust} partitions at different scales of granularity (different resolutions). Robust partitions are defined to be partitions that are (1) persistent across scale values and (2) reproducible by the deployed optimizer (e.g., Louvain) for its particular scale \citep{lambiotte2014random}. The multi-scale Markov stability for community detection is defined \citep{Schaub2019} as the following graph optimization problem: Given graph $G$, find a series of partitions at different values of the scale parameter $t$ by maximizing the \textit{generalized Markov Stability} function in Eq.\ \eqref{eq:gMS}.
\begin{equation}
 \label{eq:gMS}
 H^*(t)=\underset{H}{\operatorname{argmax}} Q_{g e n}(t, H):=\underset{H}{\operatorname{argmax}} \operatorname{Tr}\left[H^T\left(F(t)-\sum_{k=1}^m v_{2 k-1} v_{2 k}^T\right) H\right]
\end{equation}

In Eq. \eqref{eq:gMS}, the output, $H^*(t)$, is a series of $n \times c$ matrices, where $n$ represents the number of nodes in $G$, and $c$ represents the number of communities in the optimized partition of the $n$ nodes, based on the scale parameter $t$. $F(t)$ is an $n \times n$ node similarity matrix that depends on $t$. Moreover, $\left\{v_k\right\}_{k=1}^{2m}$ is a set of $n$-dimensional vector pairs that encode a null model of rank $m$. The null model is the baseline upon which the objective function of the partition is calculated. The scale parameter $t$ (also known as the Markov scale) controls the resolution of the partition $H^*(t)$. 

\subsection{The PyGenStability algorithm}
\label{ss:pygen}
The PyGenStability algorithm \citep{pygenstability_2023} uses one of the two heuristic optimizers Louvain\footnote{We use the default optimizer of PyGenStability which is the Louvain algorithm.} \citep{blondel_fast_2008} and Leiden \citep{traag_louvain_2019} to approach the maximums of the objective function in Eq. \eqref{eq:gMS} for a range of values of $t>0$ that spans from the finest to the coarsest resolution\footnote{We use the default parameters of PyGenStability including the default scale range of $-2.0$ to $0.5$ as recommended in the PyGenStability documentation.}. Among the handful of constructors that are available in the PyGenStability library, we use the \textit{continuous normalized Markov stability}: $\text{exp}(-t\textbf{L})$. This constructor is based on normalized graph Laplacian $\textbf{L}=\textbf{D}-\textbf{A}$ \citep{pygenstability_2023} where $\textbf{D}$ is the diagonal degree matrix of the input graph $G$ and $\textbf{A}$ is the adjacency matrix of graph $G$. It has the node similarity matrix function $F(t) = t\frac{\textbf{A}}{2m}$ and its null model is based on ${(v_{0})}_i={(v_{1})}_i=\frac{d_i}{2m}$.

\subsection{The heuristic selection of robust partitions}
\label{ss:heuristic}
PyGenStability produces a long sequence of partitions among which some partitions are robust. The heuristic post-processing feature of PyGenStability \citep{schindler2023multiscale} can be used to obtain those fewer robust partitions at different scales of varying coarseness. This post-processing step \citep{schindler2023multiscale} selects robust partition by relying on Normalized Variation of Information (NVI) to compare different partitions obtained for each range of the scale parameter $t$. The persistence across scales (the first characteristic from the definition of a robust partition \citep{lambiotte2014random}) is operationalized by computing the pairwise NVI for partitions across different scales and finding regions that indicate high persistence across scales \citep{schindler2023multiscale}. For each Markov scale $t$, the reproducibility to the optimization (the second characteristic from the definition of a robust partition \citep{lambiotte2014random}) is operationalized by repeating the Louvain optimization algorithm 300 times with different random initializations and computing the average pairwise NVI for the resulting ensemble of partitions. The NVI is then used to quantify the extent of reproducibility for scale value $t$ \citep{schindler2023multiscale}. Additional technical details about PyGenStability are available in \citep{pygenstability_2023}.

\subsection{Partition evaluation measures}

We use two measures of partition similarity to assess the similarity between the partition produced by a CD algorithm and a reference partition (e.g., the planted partition of a synthetic benchmark network).

The first measure is the \textit{adjusted mutual information} \citep{vinh_AMI}. It is adjusted for the fact that the mutual information is generally higher for two partitions with a larger number of communities \citep{vinh_AMI}. Adjusted mutual information has been established as a reliable measure of partition similarity \citep{jerdee2023normalized}. However, a challenge with adjusted mutual information is that its range depends on the input node set. Therefore, it is useful \citep{jerdee2023normalized} to normalize it so that extreme similarity and extreme dissimilarity (between two partitions of a network) are reflected by the numbers 1 and 0 respectively regardless of the input node set. 

Consider that $I(P_1,P_2)$ denotes the adjusted mutual information between the two partitions $P_1$ and $P_2$ for the graph $G=(V,E)$. Jerdee and Newman discuss several ways of obtaining a normalized adjusted mutual information and their implications \citep{jerdee2023normalized}. Following their results, we symmetrically normalize adjusted mutual information as in Eq.\ \eqref{eq1}.

\begin{equation}
 \label{eq1}
 I^{(S)}{(P_1,P_2)}=\frac{I(P_1,P_2)}{1/2[I(P_1,P_1)+I(P_2,P_2)]}
\end{equation}

$I^{(S)}{(P_1,P_2)}$ denotes the \textit{symmetric normalization of adjusted mutual information} (AMI for short) \citep{vinh_AMI} and quantifies the similarity between two partitions $P_1$ and $P_2$ for graph $G$. Unlike $I(P_1,P_2)$ whose value depends on the input graph, the values for $I^{(S)}{(P_1,P_2)}$ can be compared or aggregated across different networks because its expectation ranges in the unit interval and does not depend on the input graph.

The second measure that we use is the \textit{element-centric similarity} (ECS) \citep{gates2019element}. ECS quantifies the similarity between two partitions based on the common node membership in the partition as opposed to overlaps between clusters \citep{gates2019element}. We use ECS\footnote{For computing ECS, we use the default value of $0.9$ for its $\alpha$ parameter as suggested in \citep{gates2019element} and used in the documentation of the CluSim Python library.} to complement AMI results and comparisons. ECS offers several methodological advantages \citep{gates2019element} compared to most other partition similarity metrics including the widely used and problematic \citep{jerdee2023normalized,mahmoudi2024proofNMI} normalized mutual information (NMI) \citep{danon2005comparing}. 

For two identical partitions, AMI and ECS yield a value of 1. Conversely, for two highly dissimilar partitions, AMI and ECS both return small values close to 0. AMI sometimes takes negative values close to zero, while ECS is always non-negative.

\section{Machine Learning Based Scale Selection}
\label{s:ml}
The scale parameter $t$ impacts the Markov stability objective function according to Eq.\ \eqref{eq:gMS} \citep{pygenstability_2023}. We consider $t$ as a learnable parameter and develop a simple pre-trained supervised learning model for selecting the suitable scale parameter for any input graph within the PO algorithm. The development of the pre-trained scale selection model is explained in Sections \ref{ss:generate}--\ref{ss:model}. 

\subsection{Generating benchmark graphs for supervised learning}
\label{ss:generate}

For the predictive modeling task of learning the scale parameter, we develop a labeled dataset of $10^4$ synthetic graphs with planted communities using the \textit{Artificial Benchmarks for {Community} {Detection}} (ABCD) model \citep{kaminski_artificial_2021}. ABCD is the more recent alternative to the \textit{Lancichinetti-Fortunato-Radicchi} (LFR) model \citep{lancichinetti_benchmark_2008} for generating benchmark graphs that have planted communities. ABCD offers additional benefits compared to LFR such as higher scalability and a more controllable mixing parameter \citep{kaminski_artificial_2021}. The mixing parameter of the ABCD model is denoted by $\xi\in[0,1]$, and it controls the association between the structure and the communities. Small (large) values of $\xi$ lead to benchmark graphs whose planted partitions are generally easy (hard) to retrieve for CD algorithms. 

The $10^4$ graphs are generated after setting a unique random number seed and using the following parameters: the number of nodes ($n$) is randomly selected from the range of $[100, 1000)$; the minimum degree $d_{min}$ and minimum community size $k_{min}$ are randomly selected from the ranges $[1, n/{\text{divisor}}_{d})$ and $[1, n/{\text{divisor}}_{k})$ respectively, where ${\text{divisor}}_{d}$ and ${\text{divisor}}_{k}$ are both randomly selected from the range $[2, 50)$; the maximum community size is randomly selected from the range $[k_{min} + 1, n)$; the maximum degree is randomly selected from $[d_{min} + 1, n)$; and the power law exponents for the both distributions of community size and the degrees are randomly selected from the interval $(1, 16)$ and then rounded to two decimal places. Furthermore, the mixing parameter $\xi$ is selected uniformly at random from the range $[0.01, 0.5]$.

\subsection{Feature engineering for supervised learning}
\label{ss:feature}

We use the Graph2vec \citep{narayanangraph2vec} algorithm to embed each of the $10^4$ benchmark graphs into a vector with 256 dimensions. For each graph, the entries of the embedding vector are then summarized using five summary statistics: maximum, minimum, mean, median, and standard deviation. The feature set also includes six graph-level measures: number of nodes, number of edges, global efficiency, average degree, average clustering coefficient, and assortativity. In total, there are 11 features (summarizing the graph structure) and one response for this supervised learning task. The response variable for each graph used in the training dataset is $t^*$ which is the scale parameter corresponding to the partition from PyGenStability which has the highest AMI (among the multiple partitions produced by PyGenStability) with the planted partition.

\subsection{Preparing the labeled dataset for supervised learning}
\label{ss:labeled}

To obtain the response (target) variable for each graph in the training set, we run the PyGenStability algorithm (followed by the robust post-processing heuristic \citep{schindler2023multiscale}, described in \ref{ss:heuristic}) to obtain several partitions each corresponding to a value for the scale parameter $t$. For each value of $t$, we then calculate an AMI by comparing the corresponding partition with the planted partition. Finally, as the response value for each graph, we use $t^*$ which is the scale parameter corresponding to the partition that has the largest AMI value for that input graph. The labeled dataset for supervised learning is therefore made up of 11 predictor variables (features) and one response variable for each of the $10^4$ benchmark graphs. The next step is the standard task of fitting a supervised Machine Learning (ML) model on this tabular labeled dataset.

\subsection{Model selection for supervised learning}
\label{ss:model}
The labeled dataset is randomly split into accessible data (80\%) and hold-out data (20\%). We then use the accessible data to train eight predictive models. The eight predictive model types considered are ridge regression, lasso regression, decision trees, random forests, gradient boosting, histogram-based gradient boosting, elastic net, and support vector regression. For the hyperparameter tuning of each of the eight models, we use a randomized search cross-validation with 5 folds and use the default scoring method of each model as the loss function. Through this process, we construct eight tuned supervised learning models (one for each model type). 

Among these eight tuned models, we select the model that has the lowest mean squared error (MSE) and mean absolute error (MAE) on the hold-out set. Table~\ref{tab:mse} shows the MSE and MAE of each of the tuned predictive models on the hold-out test set. The MSE and MAE of the gradient boosting model on the hold-out set are 0.395 and 0.520 respectively which are the lowest error rates among the eight tuned models considered. The tuned gradient boosting model has the following hyperparameter setting: number of trees: 141, learning rate: 0.027, maximum depth: 6, and subsampling parameter: 0.790. Given the results in Table~\ref{tab:mse}, we choose the tuned gradient boosting model with the above-mentioned hyperparameters as the selected model for predicting the scale parameter within the PO algorithm. 

\begin{table}[htbp!]
 \centering
 \caption{Mean Absolute Error (MAE) and Mean Squared Error (MSE) for each trained and tuned predictive model on the hold-out test set}
 \begin{tabular}{ccc} \hline 
 Type of predictive model& MAE& MSE\\ \hline 
 Decision Tree& 0.418& 0.536\\ 
 Random Forest& 0.398& 0.520\\ 
 Gradient Boosting& 0.395& 0.520\\ 
 Histogram-based Gradient Boosting& 0.402& 0.521\\
 Ridge Regression& 0.483& 0.590\\ 
 Lasso Regression& 0.531& 0.644\\ 
 Elastic Net& 0.490& 0.598\\ 
 Support Vector Regression& 0.510& 0.582\\ 
 \hline
 \end{tabular}
 \label{tab:mse}
\end{table}

After both steps of model tuning and model selection are complete, the overall best model and its details are determined: gradient boosting model and its tuned hyperparameters. At this stage, we retrain a fresh gradient boosting model with the same hyperparameter setting on all the $10^4$ samples of the labeled dataset and save it as a pre-trained predictive model for the task of predicting the scale parameter $t$ for any input graph within the PO algorithm. This model (as a part of PO) will be used on the downstream task of retrieving planted communities in a fresh set of benchmark graphs as well as real networks (with results reported in Section \ref{s:result}). At this stage, the practical relevance of using our proposed ML-based scale selection remains unknown. The comparative performance results of PO with respect to other CD algorithms will partly show the extent to which our proposed idea of using this ML-based scale selection is practical and useful.

\section{Results}
\label{s:result}

Now that the PO algorithm including its pre-trained scale selection model is ready, we use real and synthetic networks to compare PO with the 29 other CD algorithms (reviewed in Section \ref{s:review}). 

To access the 29 existing CD algorithms, we use their publicly available implementations from four Python libraries. Specifically, for the two algorithms MR and MV, we utilize the \textit{PyGenStability} library (version 0.2.2) \citep{pygenstability_2023}. For the four algorithms SB, SM, BP, and BM, we rely on the \textit{graph\_tool} library (version 2.57) \citep{graph-tool_2014}. As for asynchronous label propagation and the Kernighan-Lin algorithms, we turn to the \textit{NetworkX} library (version 3.1) \citep{networkx}. For the other 21 algorithms, we rely on the Community Discovery library (\textit{CDlib}) (version 0.2.6) \citep{rossetti2019cdlib}. The documentations of these four libraries and the main articles indicate no instructions or procedures for tuning any hyperparameters before usage. Given this and in the interest of representing the most common usage, we use all the 29 algorithms without making any changes to their default settings. 

\subsection{Qualitative assessment of partitions from each family of algorithms}
\label{ss:visual}

\begin{figure}[!htbp]
     \centering
          \begin{subfigure}[b]{0.485\textwidth}
         \centering
         \includegraphics[trim={3cm 3cm 3cm 3cm},clip,width=\textwidth]{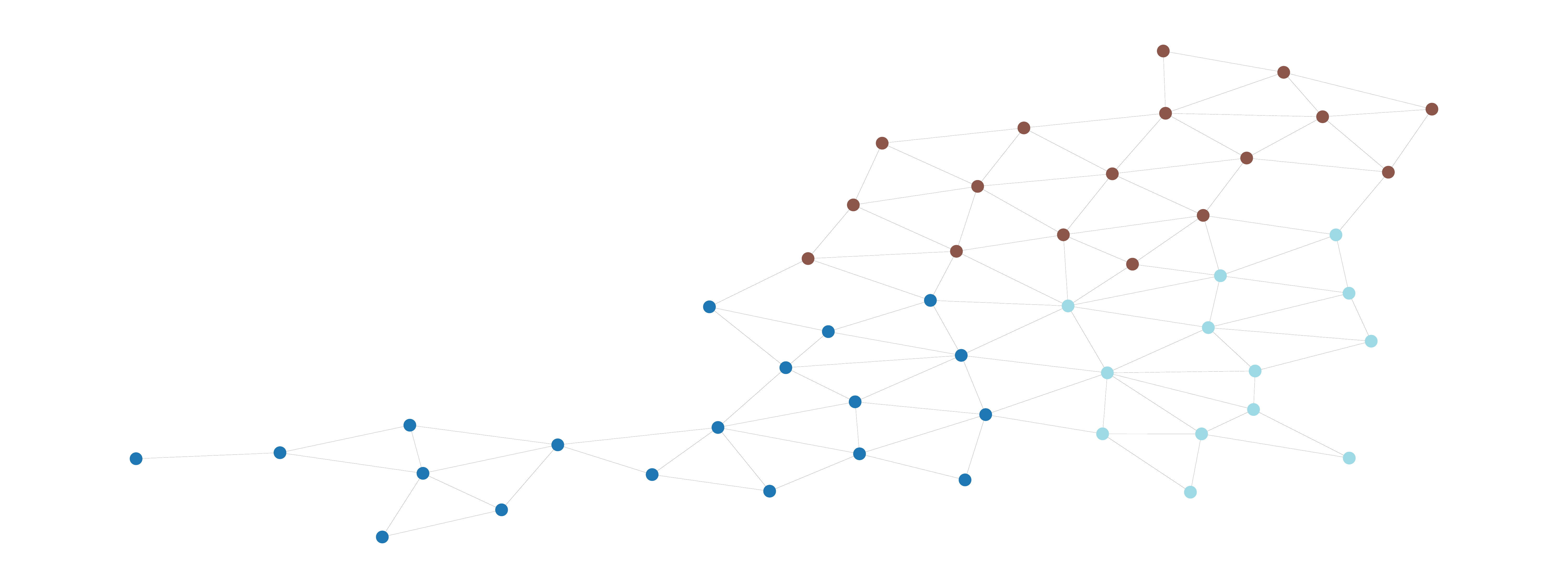}
         \caption{Paris}
         \label{subfig:paris30}
     \end{subfigure}
    \begin{subfigure}[b]{0.485\textwidth}
         \centering
         \includegraphics[trim={3cm 3cm 3cm 3cm},clip,width=\textwidth]{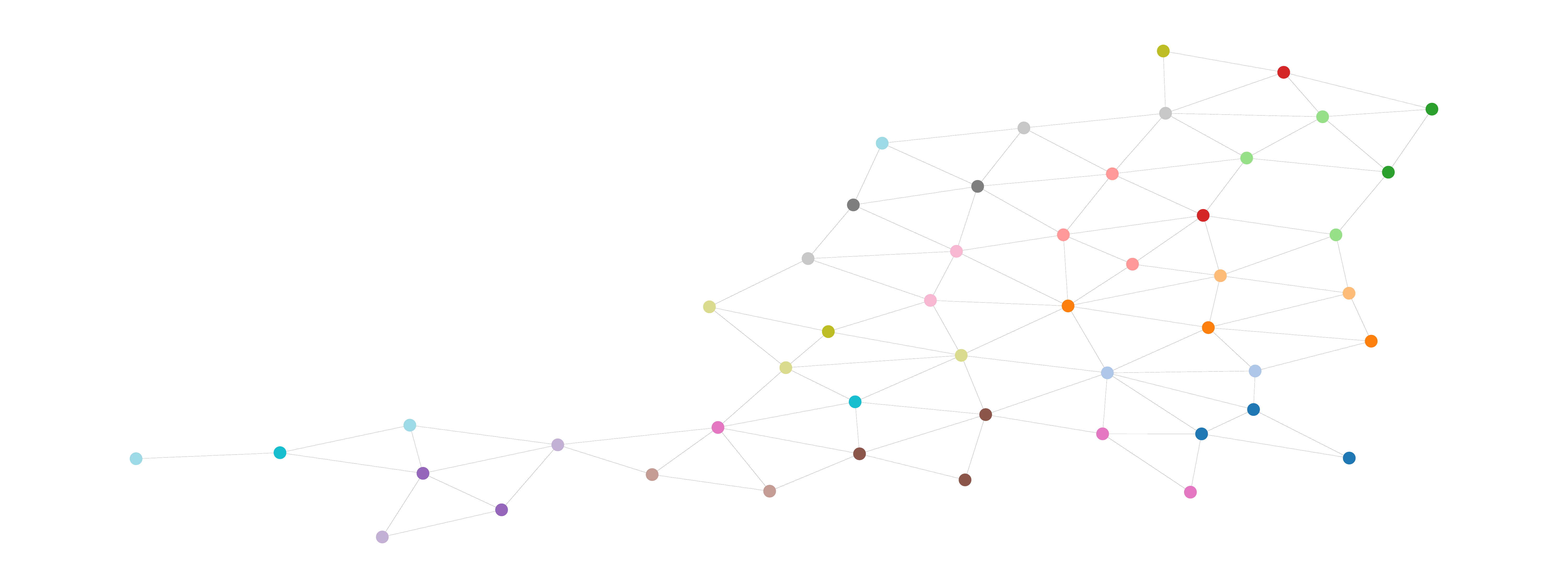}
         \caption{Constant Potts Model (CPM)}
         \label{subfig:cpm30}
     \end{subfigure}
     \begin{subfigure}[b]{0.485\textwidth}
         \centering
         \includegraphics[trim={3cm 3cm 3cm 3cm},clip,width=\textwidth]{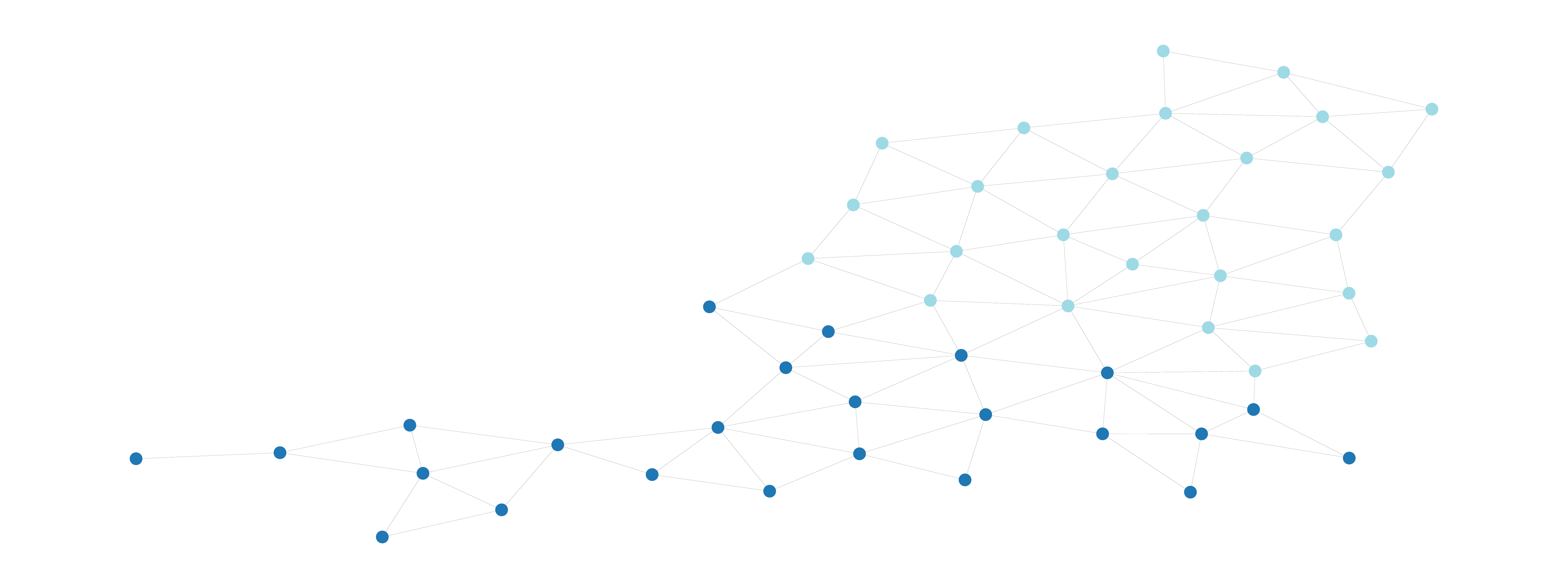}
         \caption{Stochastic Block Model (SBM)}
         \label{subfig:sbm30}
     \end{subfigure}
    \begin{subfigure}[b]{0.485\textwidth}
         \centering
         \includegraphics[trim={3cm 3cm 3cm 3cm},clip,width=\textwidth]{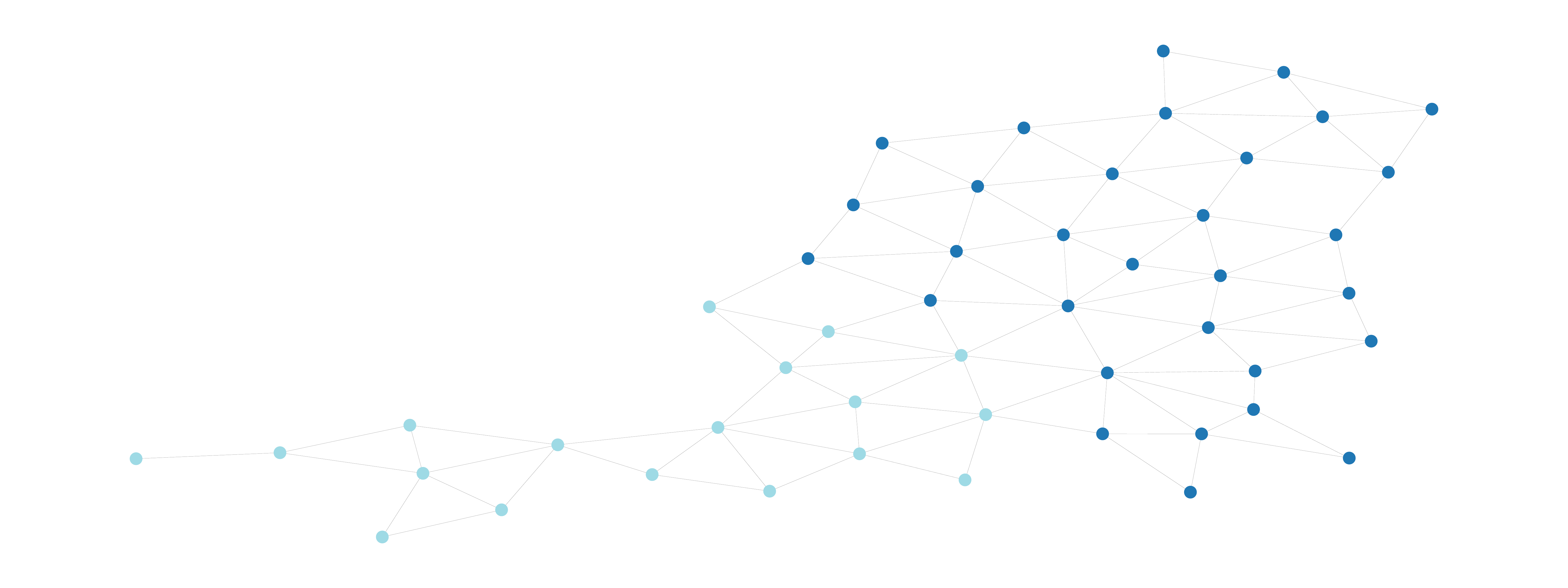}
         \caption{Bayesian Planted Partition (BPP)}
         \label{subfig:bpp30}
     \end{subfigure}
     \begin{subfigure}[b]{0.485\textwidth}
         \centering
         \includegraphics[trim={3cm 3cm 3cm 3cm},clip,width=\textwidth]{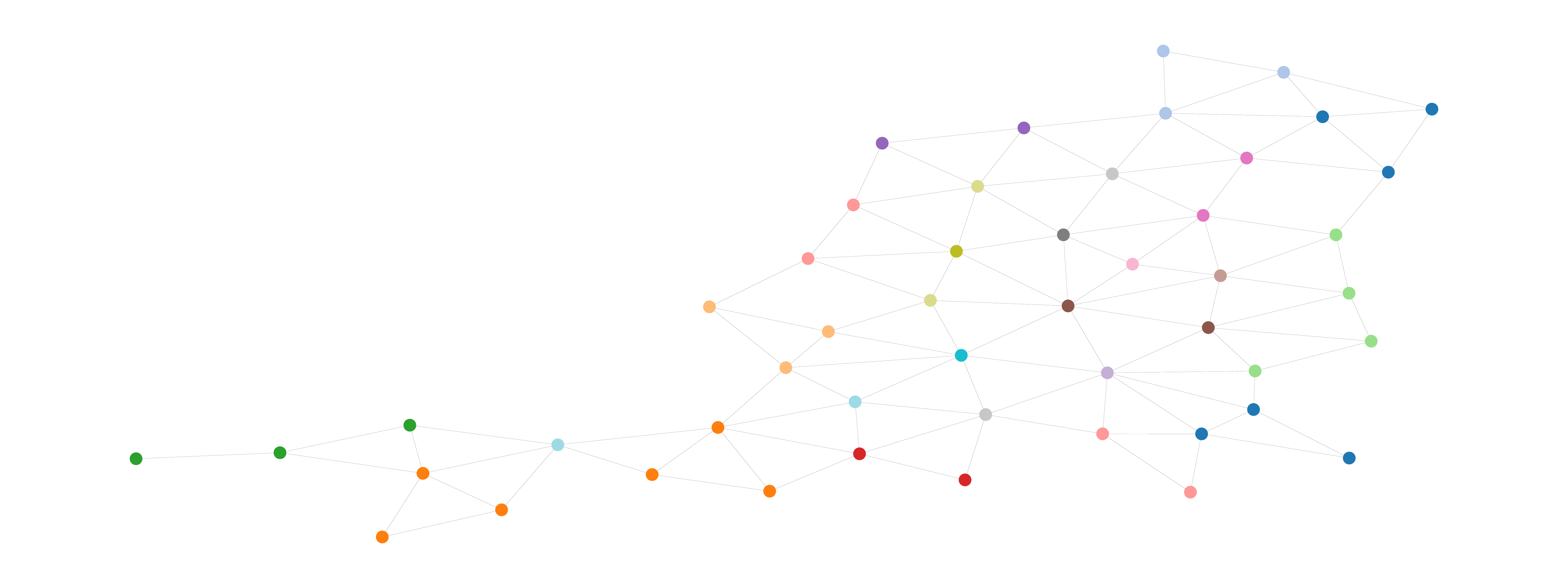}
         \caption{Markov stability with random partition selection (MR)}
         \label{subfig:mr30}
     \end{subfigure}
     \begin{subfigure}[b]{0.485\textwidth}
         \centering
         \includegraphics[trim={3cm 3cm 3cm 3cm},clip,width=\textwidth]{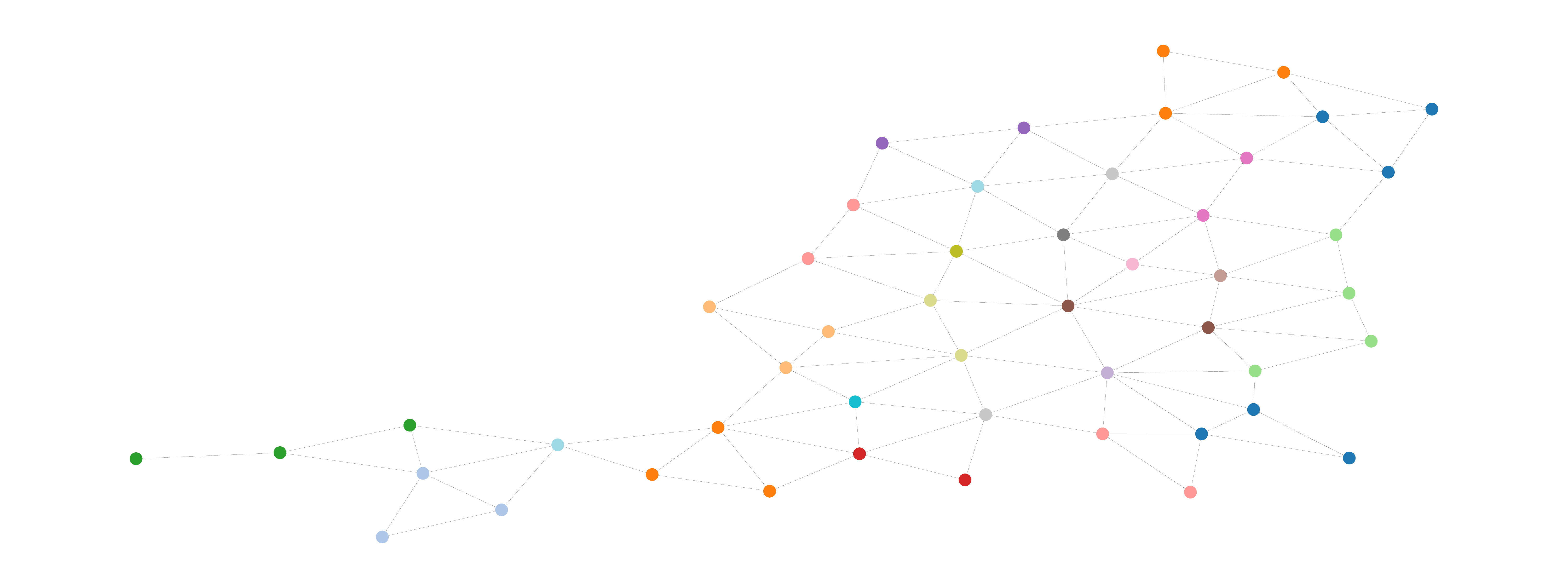}
         \caption{Markov Stability with minimum NVI (MV)}
         \label{subfig:mv30}
     \end{subfigure}
     \begin{subfigure}[b]{0.485\textwidth}
         \centering
         \includegraphics[trim={3cm 3cm 3cm 3cm},clip,width=\textwidth]{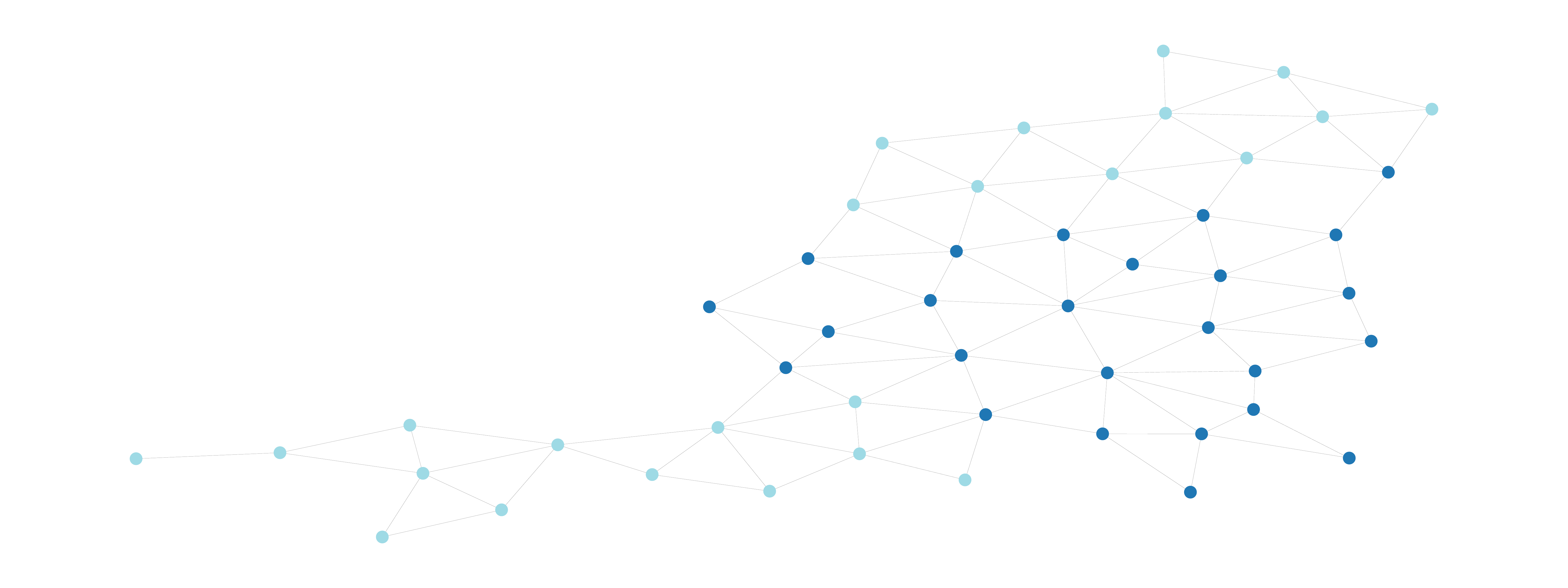}
         \caption{Diffusion Entropy Reducer (DER)}
         \label{subfig:der30}
     \end{subfigure}
     \begin{subfigure}[b]{0.485\textwidth}
         \centering
         \includegraphics[trim={3cm 3cm 3cm 3cm},clip,width=\textwidth]{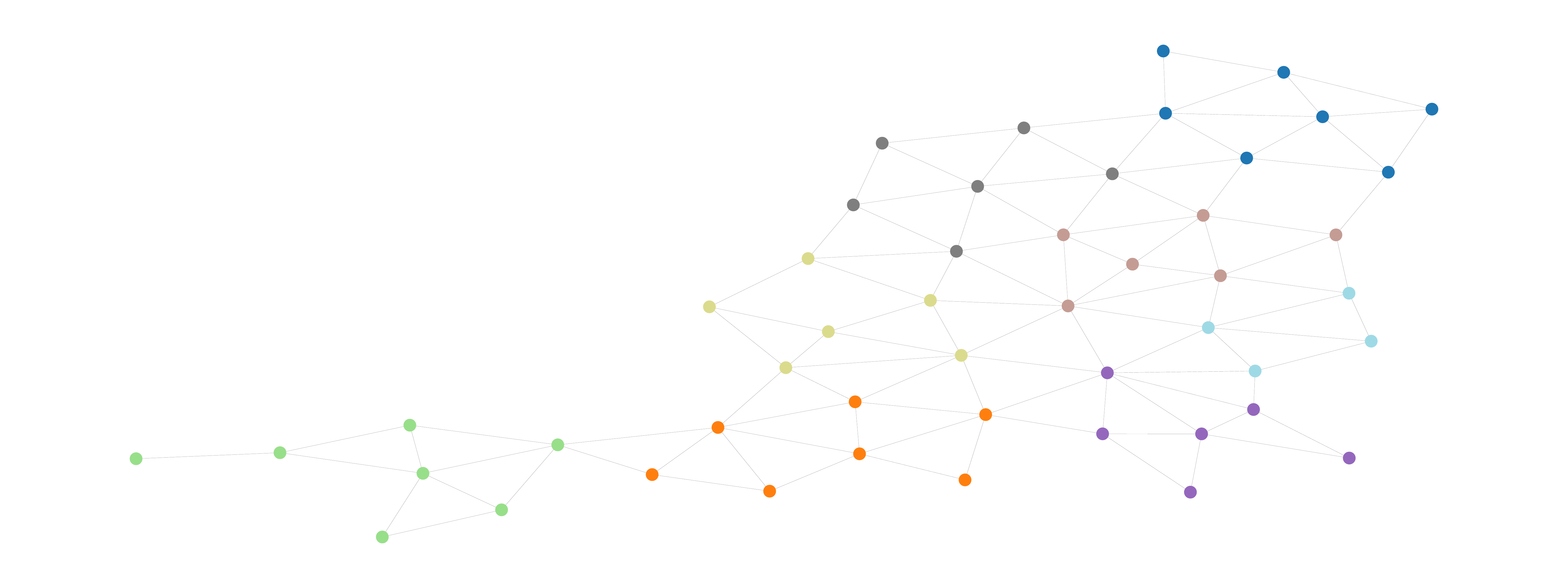}
         \caption{The proposed algorithm: PyGenStabilityOne (PO)}
         \label{subfig:po30}
     \end{subfigure}
        \begin{subfigure}[b]{0.485\textwidth}
         \centering
         \includegraphics[trim={3cm 3cm 3cm 3cm},clip,width=\textwidth]{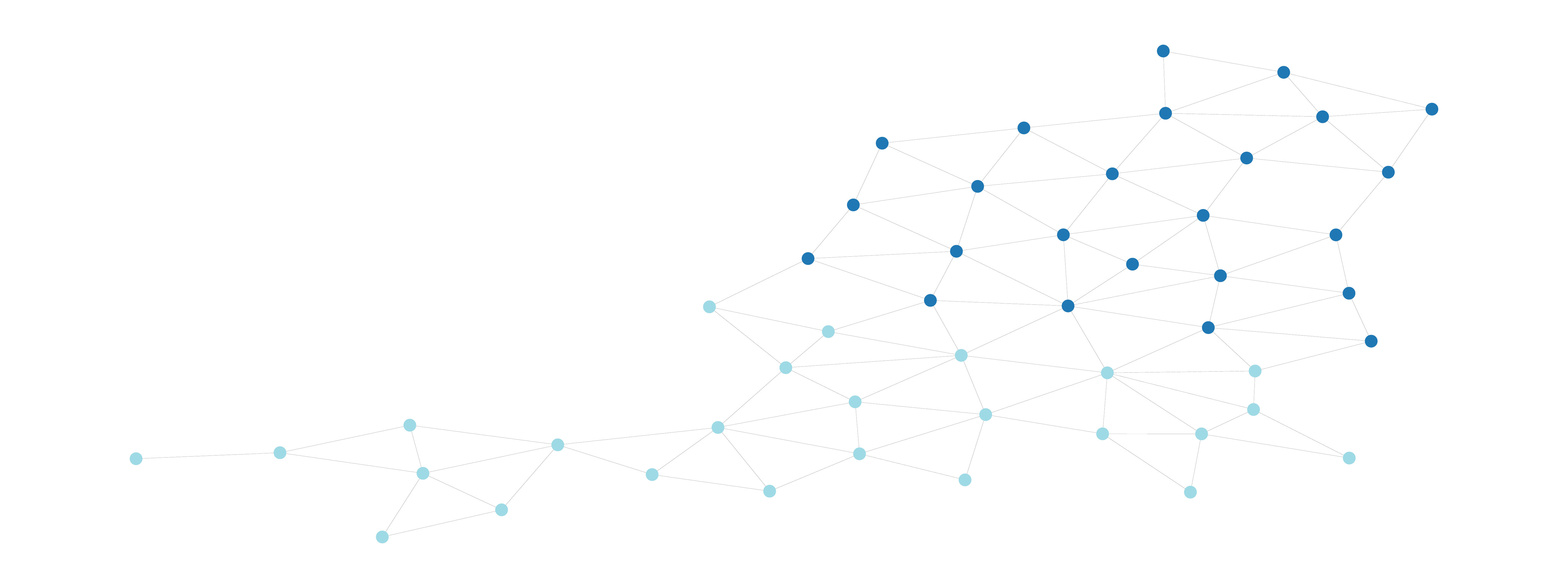}
         \caption{Kernighan Lin}
         \label{subfig:kl30}
     \end{subfigure}
     \begin{subfigure}[b]{0.485\textwidth}
         \centering
         \includegraphics[trim={3cm 3cm 3cm 3cm},clip,width=\textwidth]{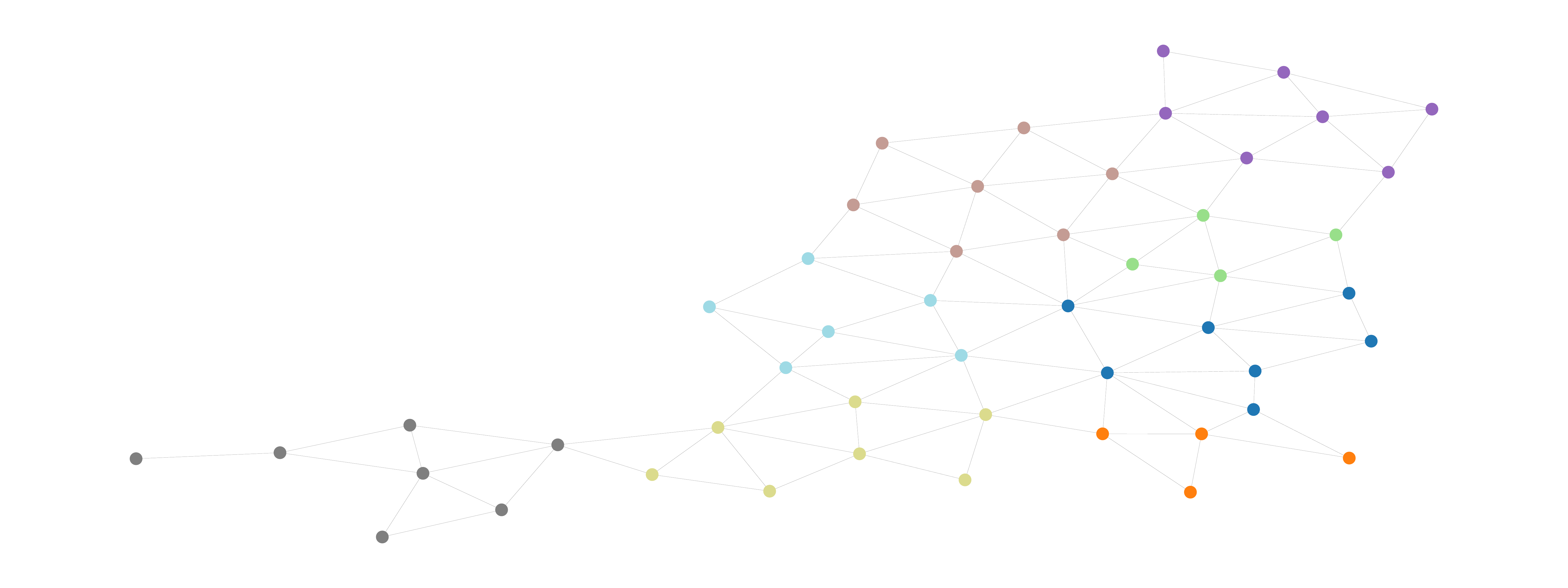}
         \caption{Asynchronous label propagation}
         \label{subfig:alp30}
     \end{subfigure}
        \caption{Community detection for the Contiguous USA network using ten methods leading to ten different partitions as shown by node colors (panels a-j). (Magnify the high-resolution color figure on screen for more details.) }
        \label{fig:contiguous}
        \Description[<short description>]{<long description>}  
\end{figure}

\begin{figure}[!htbp]
     \centering
          \begin{subfigure}[b]{0.485\textwidth}
         \centering
         \includegraphics[trim={3cm 3cm 3cm 3cm},clip,width=\textwidth]{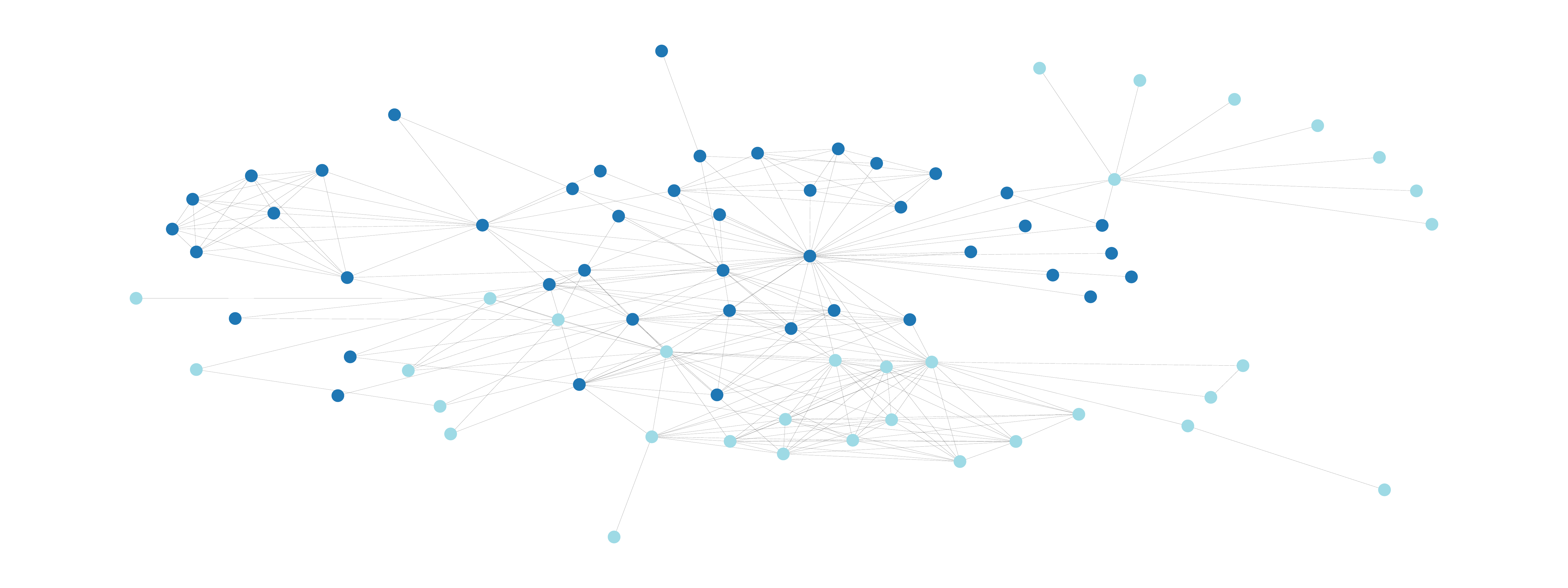}
         \caption{Paris}
         \label{subfig:paris40}
     \end{subfigure}
    \begin{subfigure}[b]{0.485\textwidth}
         \centering
         \includegraphics[trim={3cm 3cm 3cm 3cm},clip,width=\textwidth]{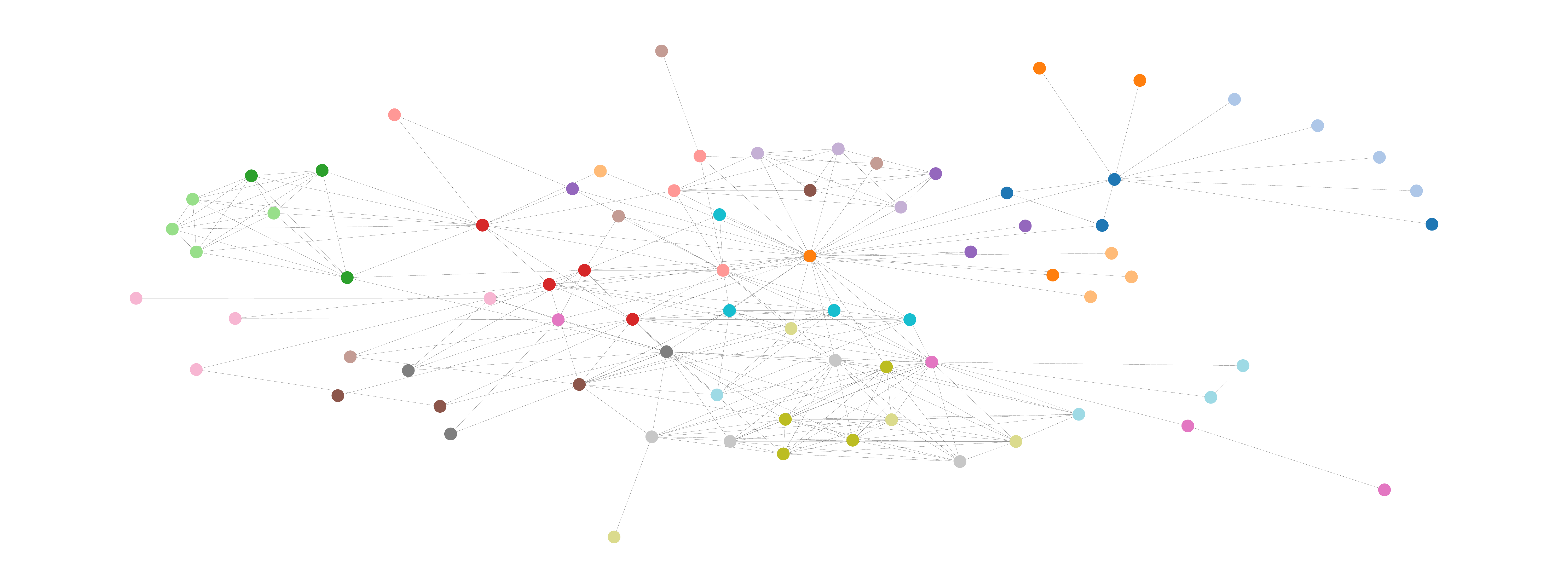}
         \caption{Constant Potts Model (CPM)}
         \label{subfig:cpm40}
     \end{subfigure}
     \begin{subfigure}[b]{0.485\textwidth}
         \centering
         \includegraphics[trim={3cm 3cm 3cm 3cm},clip,width=\textwidth]{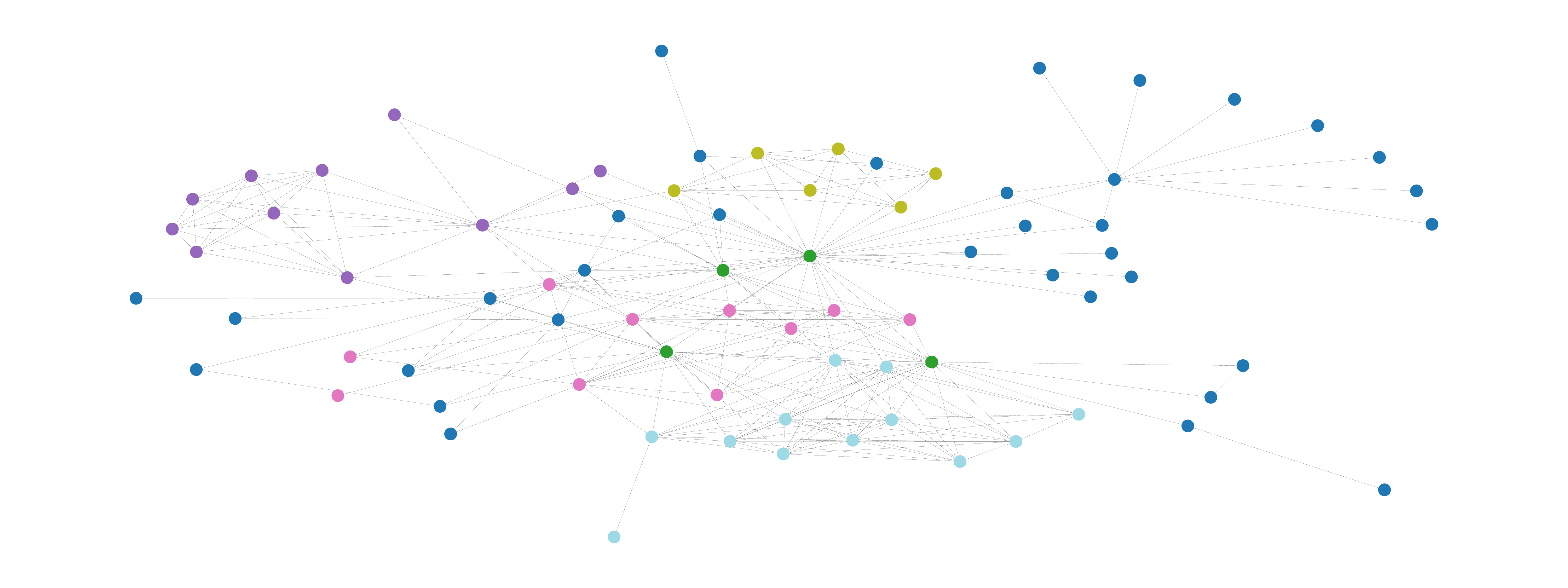}
         \caption{Stochastic Block Model (SBM)}
         \label{subfig:sbm40}
     \end{subfigure}
    \begin{subfigure}[b]{0.485\textwidth}
         \centering
         \includegraphics[trim={3cm 3cm 3cm 3cm},clip,width=\textwidth]{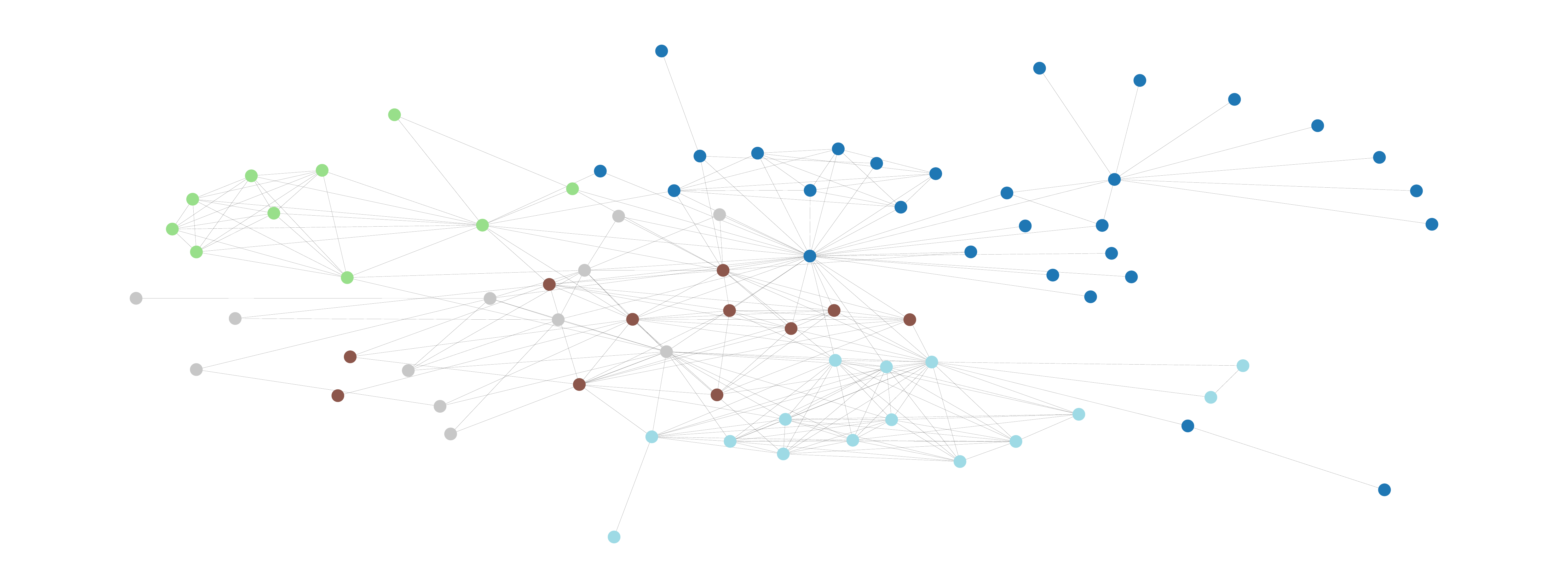}
         \caption{Bayesian Planted Partition (BPP)}
         \label{subfig:bpp40}
     \end{subfigure}
     \begin{subfigure}[b]{0.485\textwidth}
         \centering
         \includegraphics[trim={3cm 3cm 3cm 3cm},clip,width=\textwidth]{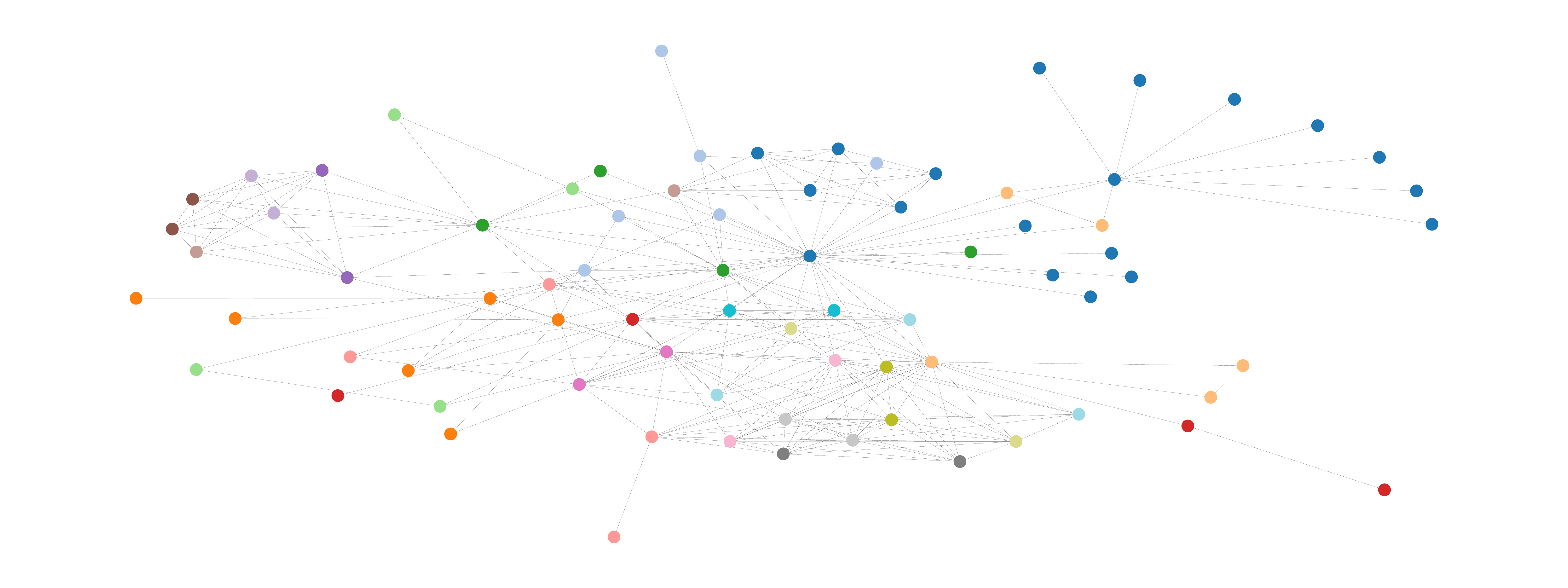}
         \caption{Markov stability with random partition selection (MR)}
         \label{subfig:mr40}
     \end{subfigure}
     \begin{subfigure}[b]{0.485\textwidth}
         \centering
         \includegraphics[trim={3cm 3cm 3cm 3cm},clip,width=\textwidth]{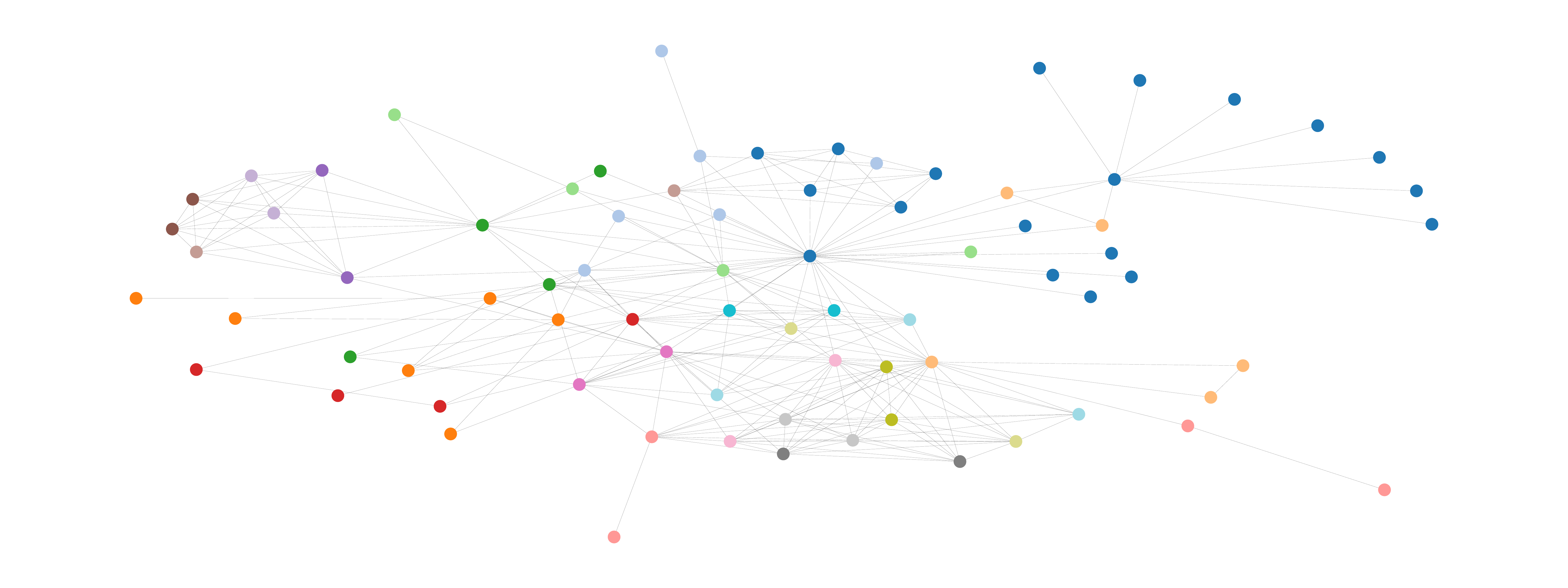}
         \caption{Markov stability with minimum NVI (MV)}
         \label{subfig:mv40}
     \end{subfigure}
     \begin{subfigure}[b]{0.485\textwidth}
         \centering
         \includegraphics[trim={3cm 3cm 3cm 3cm},clip,width=\textwidth]{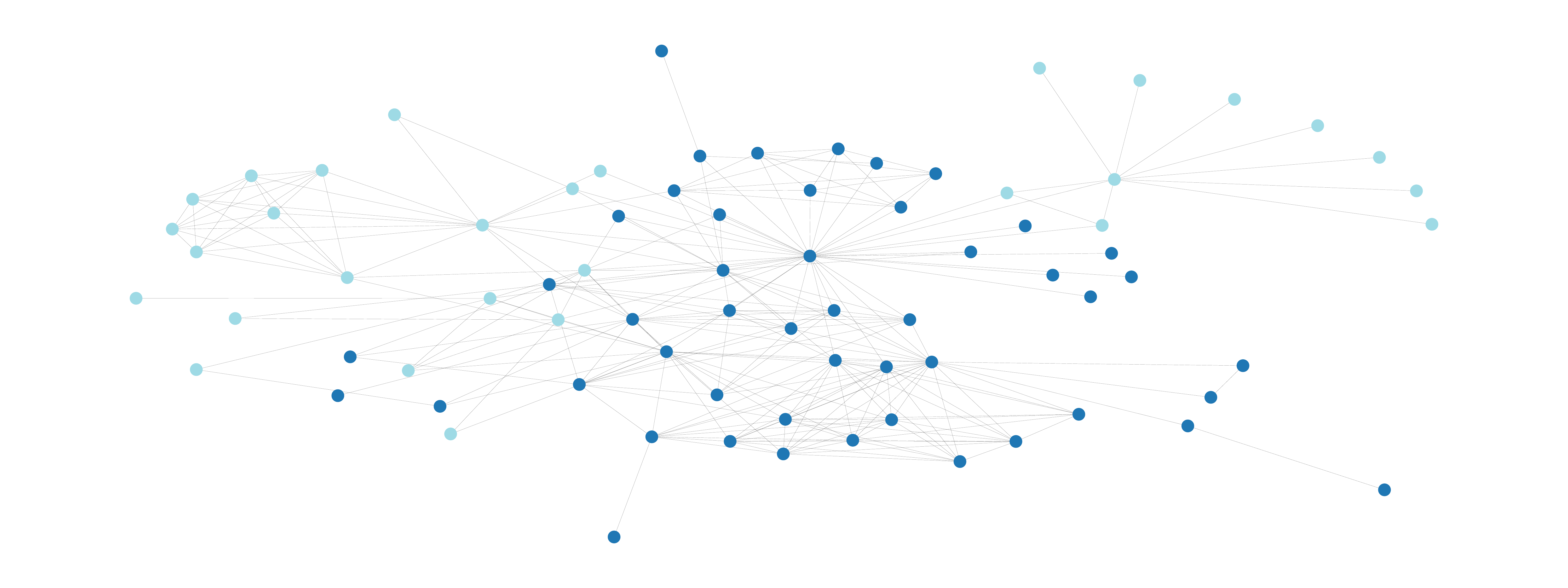}
         \caption{Diffusion Entropy Reducer (DER)}
         \label{subfig:der40}
     \end{subfigure}
     \begin{subfigure}[b]{0.485\textwidth}
         \centering
         \includegraphics[trim={3cm 3cm 3cm 3cm},clip,width=\textwidth]{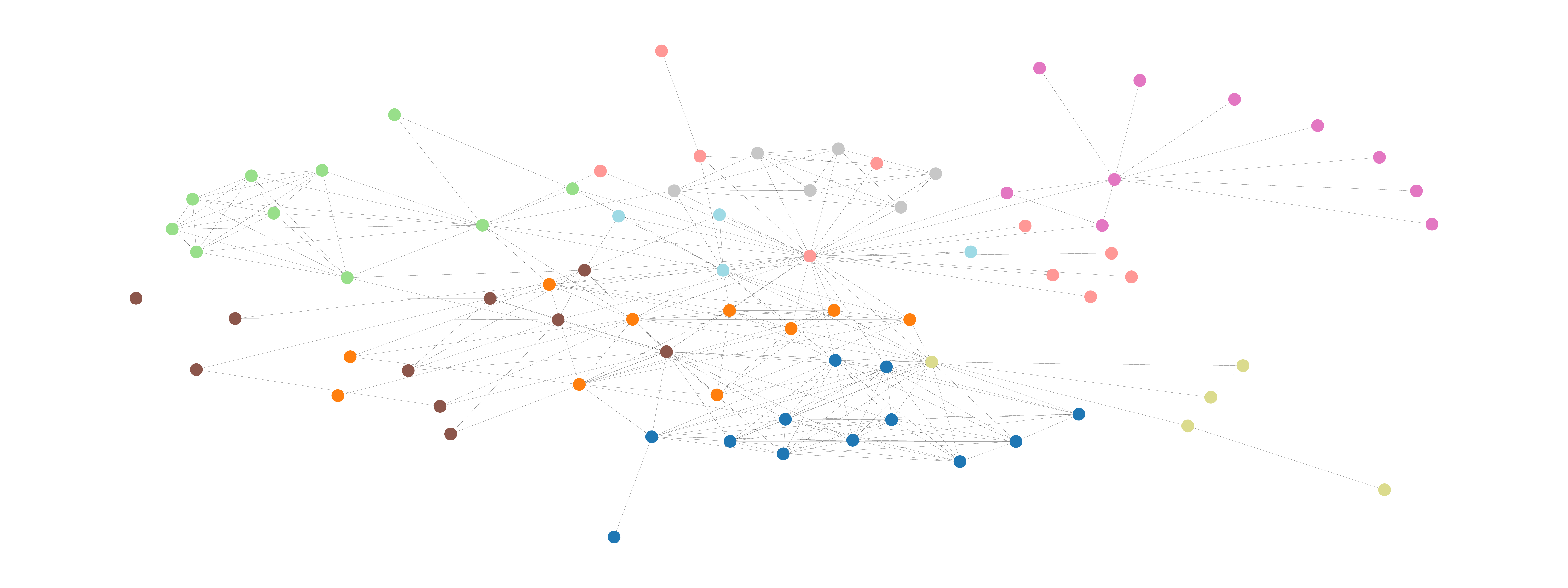}
         \caption{The proposed algorithm: PyGenStabilityOne (PO)}
         \label{subfig:po40}
     \end{subfigure}
        \begin{subfigure}[b]{0.485\textwidth}
         \centering
         \includegraphics[trim={3cm 3cm 3cm 3cm},clip,width=\textwidth]{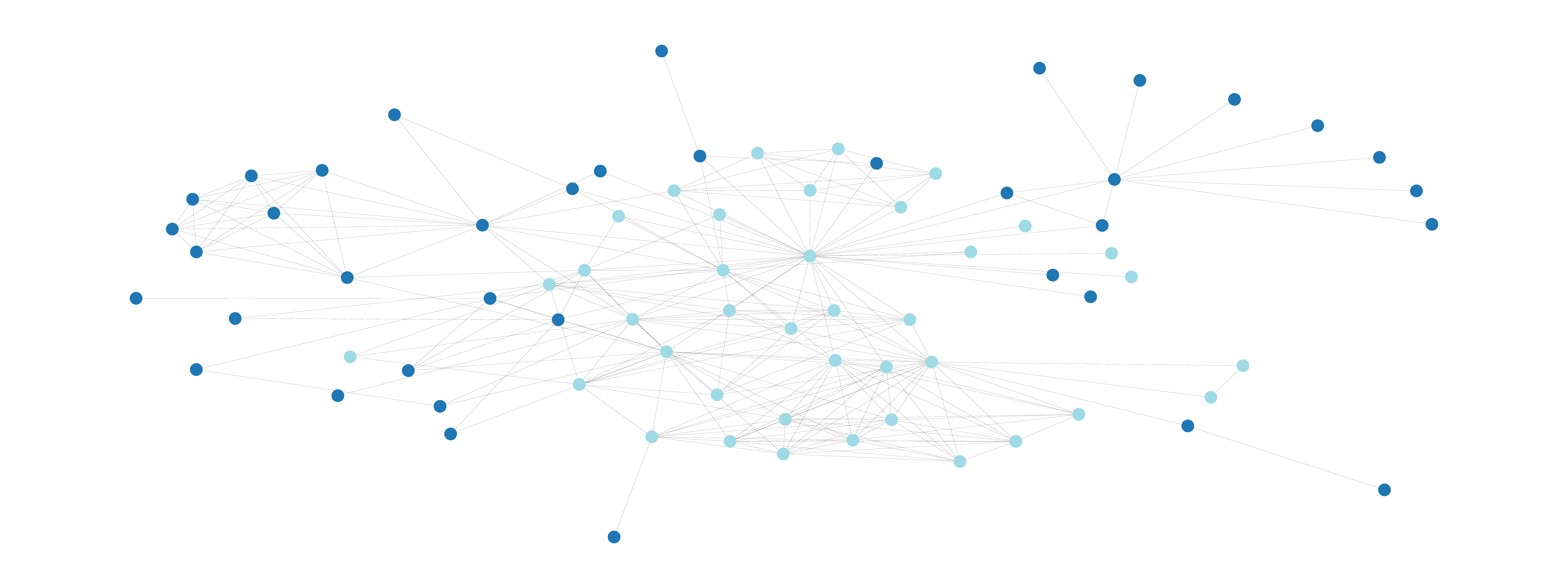}
         \caption{Kernighan Lin}
         \label{subfig:kl40}
     \end{subfigure}
     \begin{subfigure}[b]{0.485\textwidth}
         \centering
         \includegraphics[trim={3cm 3cm 3cm 3cm},clip,width=\textwidth]{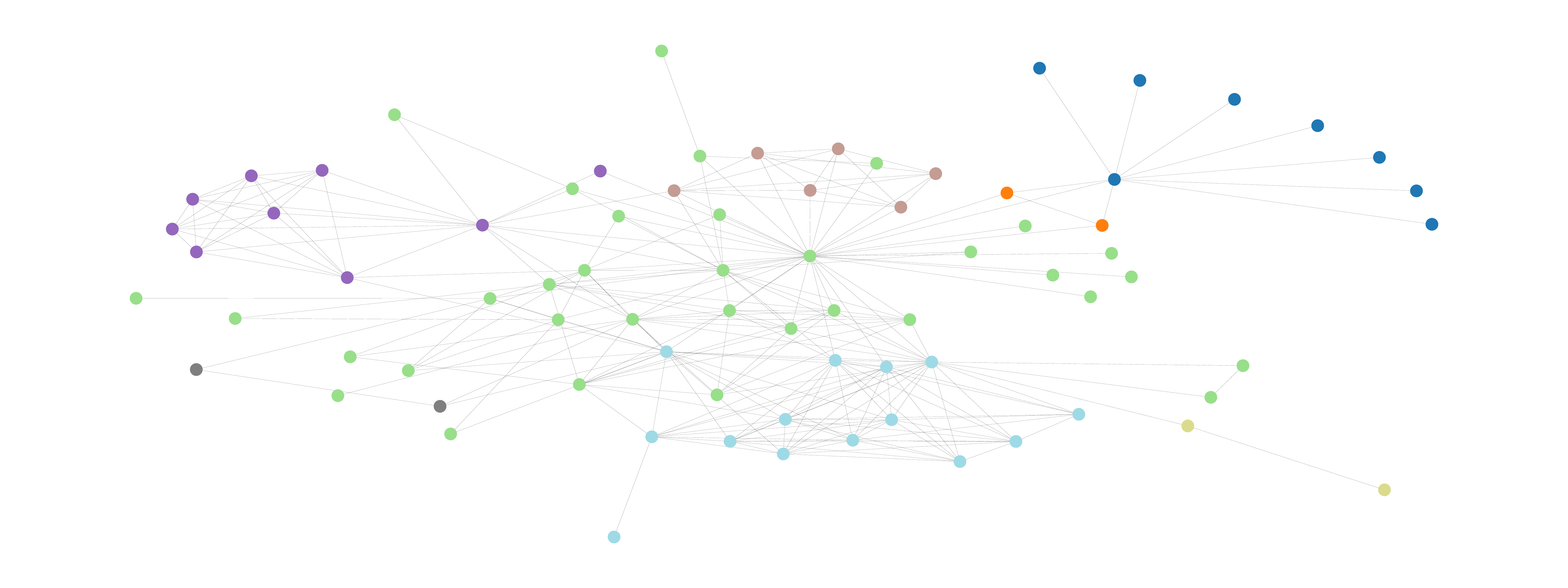}
         \caption{Asynchronous label propagation (AL)}
         \label{subfig:alp40}
     \end{subfigure}
        \caption{Community detection for the Les Mis\'erables network using ten methods leading to ten different partitions as shown by node colors (panels a-j). (Magnify the high-resolution color figure on screen for more details.) }
        \label{fig:lesmis}
        \Description[<short description>]{<long description>}
\end{figure}

In this subsection, we simply visualize the partitions of at least one algorithm from each family of the 29 reviewed algorithms on two real\footnote{The word \textit{real} is used as opposed to synthetic. We refer to networks that are not synthetically generated as real networks.} networks. The purpose is to demonstrate the fundamental differences between the output of each family and their differences to the output of PO. Both real networks used are small and unweighted, but have different structures: \textit{Contiguous USA}\footnote{The \textit{contiguous\_usa} network \citep{knuth1993stanford} has nodes representing US states and edges indicating a land-based border between two states. It is available in the \href{https://networks.skewed.de/net/contiguous_usa}{Netzschleuder} repository.}(visualized in Fig.~\ref{fig:contiguous}) is a planar network with low degree heterogeneity while \textit{Les Mis\'erables}\footnote{The \textit{lesmis} network \citep{knuth1993stanford} has nodes representing fictional characters from the novel ``Les Mis\'erables'' and edges indicating the co-appearance of a pair of characters in a scene. It is available in the \href{https://networks.skewed.de/net/lesmis}{Netzschleuder} repository.}(visualized in Fig.~\ref{fig:lesmis}) is a clustered network with a reasonably modular structure and high degree heterogeneity. 

Fig.~\ref{subfig:paris30} shows that Paris algorithm splits the Contiguous USA network into 3 communities; the dark blue community in the partition of Paris is almost disconnected (has an articulation point\footnote{A node is said to be an articulation point in a community/graph if its removal disconnects the community/graph.} in the middle). Fig.~\ref{subfig:cpm30} shows that the CPM has detected only small communities with 1-3 nodes in each; they are not particularly consistent with or interpretable by the structure, and some communities are disconnected.

While inferential and cutting methods have major differences, the partitions from the two inferential methods SBM and BPP in Figs.~\ref{subfig:sbm30}--\ref{subfig:bpp30} are particularly similar to the partition from the Kernighan-Lin bisection algorithm in Fig.~\ref{subfig:kl30}; they all partition the network into two communities with minor differences. The DER algorithm also produced two communities as shown in Fig.~\ref{subfig:der30}, but one community is positioned in the middle of two disconnected parts from the other community; the partition of DER is less interpretable than the partition of Kernighan-Lin. The two partitions from MV and MR in Figs.\ref{subfig:mr30}--\ref{subfig:mv30} are almost identical and have many communities with 1-3 nodes; some communities are disconnected. AL and PO have produced very similar partitions of 8 communities in Fig.~\ref{subfig:alp30}--\ref{subfig:po30} which seem reasonably consistent with the structure of the network. While the single example in Fig.~\ref{fig:contiguous} is not meant to cover all the advantages and disadvantages of these algorithms, it illustrates some of them for a specific planar instance. More importantly, it demonstrates the major differences between the output of these algorithms. The results from an empirical study that uses community detection in the analysis pipeline may change dramatically depending on the choice of the community detection method (we discuss this point further in Section \ref{ss:cscw} by reviewing some examples from the literature).

Fig.~\ref{subfig:paris40} shows that Paris algorithm splits the Les Mis\'erables network into two communities; the light blue community in the partition of Paris is disconnected; the dark blue community seems somewhat fragmented too. Fig.~\ref{subfig:cpm40} shows that the CPM has detected only small communities with 1-4 nodes in each; they are not particularly consistent with or interpretable by the structure, and some communities are disconnected. The two-community partitions from DER in Fig.~\ref{subfig:der40} and KL in Fig.~\ref{subfig:kl40} are particularly similar. The two partitions from MV and MR in Figs.\ref{subfig:mr40}--\ref{subfig:mv40} are almost identical again, and they both have small and medium-sized communities for the Les Mis\'erables network. AL in Fig.~\ref{subfig:alp40} has produced some communities that are explainable based on the structure, but some other communities like the orange community indicate the common limitation of propagation-based methods (that some nodes never receive the propagated labels). The two inferential methods have interesting partitions in Figs.~\ref{subfig:sbm40}--\ref{subfig:bpp40}. SBM has produced a dark blue community in Figs.~\ref{subfig:sbm40} that has many disconnected components; not only they are disconnected, some nodes of this disconnected community have a particularly long geodesic distance from each other in the graph. While SBM is claimed to have the advantage of not being restricted to assortative patterns \citep{sbm_2014}, the same feature acts as a disadvantage when the network has a modular and clustered structure that SBM fails to recover. The BPP algorithm is inferential but more grounded than SBM w.r.t.\ capturing assortative patterns. The partition from BPP in Fig.~\ref{subfig:bpp40} is easier to explain compared to the partition of SBM; yet the dark blue community remains difficult to explain particularly given the articulation point at its center. The partition from the PO algorithm is shows in Fig.~\ref{subfig:po40}; it seems to be more consistent with the structure than the other partitions in Fig.~\ref{fig:lesmis}.

The observational comparison of partitions can only take us so far. We move on to comparing the 30 algorithms systematically using synthetic benchmark networks that are particularly designed for comparing CD algorithms.

\subsection{Generating benchmark for synthetic retrieval testing of 30 algorithms}
\label{ss:generate_retrieval}

The ABCD benchmarks that we use for comparing the 30 algorithms are generated randomly so that they represent a wide range of structurally diverse networks with different mixing parameters. We generate 500 ABCD graphs using a distinct random number seed\footnote{Choosing a distinct seed prevents data leakage (i.e.\ train-test contamination) between the $10^4$ ABCD graphs used in the development of PO and these 500 ABCD graphs that are used for testing PO and other algorithms.} and the following parameters: the number of nodes ($n$) is randomly selected from the range of $[10, 1000)$; the maximum community size is randomly selected from the range $[k_{min} + 1, n)$; the maximum degree is randomly chosen from $[d_{min} + 1, n)$; the minimum degree $d_{min}$ and minimum community size $k_{min}$ are randomly chosen from the range $[1, n/4)$; and the power law exponents for both the degree distribution and the community size distribution are randomly selected from the interval $(1, 8)$ and then rounded to two decimal places. Furthermore, the parameter $\xi$ (mixing parameter) is selected from the set $\{0.01, 0.1, 0.3, 0.5, 0.7\}$ for five distinct experimental settings, each comprising 100 ABCD networks.

We run the 30 algorithms on the 500 ABCD graphs and measure the AMI and ECS of each algorithm on each graph. The AMI results are described in Section \ref{ss:abcd_AMI} and tested for statistical significance. Similarly, the ECS results on ABCD graphs are provided in Section \ref{ss:abcd_ECS}. In Section \ref{ss:abcd_summary}, ABCD results based on both AMI and ECS are summarized.

\subsection{AMI results for retrieving the planted partitions of synthetic benchmarks}
\label{ss:abcd_AMI}

\begin{figure}[htpb!]
 \includegraphics[width=\textwidth]{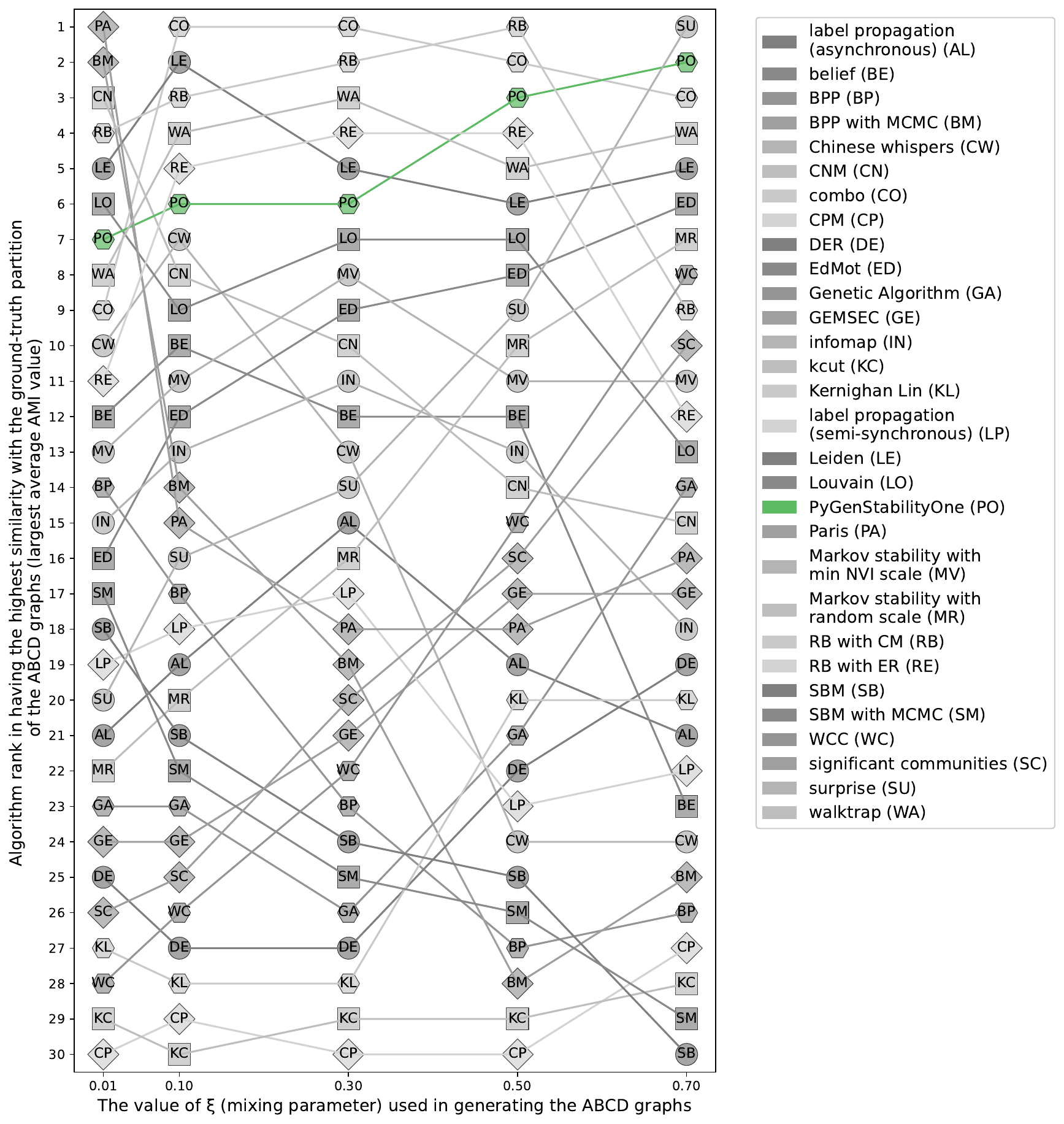}
 \caption{Ranks of 30 CD algorithms based on their average AMI for 100 ABCD graphs in five experiment settings. (Magnify the high-resolution figure on screen for details.)}
 \label{fig:Ranking_AMI}
        \Description[<short description>]{<long description>}
\end{figure}

Fig.~\ref{fig:Ranking_AMI} shows the ranking of 30 algorithms based on their average AMI for each experiment setting. While there is no algorithm that outperforms all others across all five experiment settings, PO is always among the top seven algorithms with the highest average AMI across all five experiment settings. PO and Leiden are the only two algorithms whose AMI ranks remain in the top seven across the five experiment settings. Besides PO and Leiden, there are three other algorithms which have relatively high AMI ranks across the five experiment settings: Combo, RB, and Walktrap. The ranking of these algorithms are not as stable as PO, but their ranks are always within the top nine. The two algorithms Paris and BM have remarkable performance when $\xi=0.01$, but their ranks substantially plummet for higher values of the mixing parameter $\xi$. The opposite pattern is observed for the surprise algorithm whose performance rank consistently improves as $\xi$ increases. The AMI averages of algorithms are available in Table~\ref{tab:abcdavg} in the appendix.

\begin{table}[htp!]
\centering
\caption{Statistically significant discoveries for comparing AMI values between PO and another algorithm for each value of $\xi$. * in row $a$ and column $\xi$ means that PO has a significantly higher mean AMI compared to the algorithm in row $a$ based on 100 ABCD graphs generated with the mixing parameter $\xi$. }
\begin{tabular}{lccccc}
\hline
Algorithm compared to PyGenStabilityOne & $\xi$ & $\xi$ & $\xi$ & $\xi$ & $\xi$ \\
\hline 1. Modularity-based optimization algorithms & 0.01 & 0.1 & 0.3 & 0.5 & 0.7 \\
\quad \quad Belief & & & * & * & * \\
\quad \quad Combo & & & & & \\
\quad \quad EdMot & & & & & \\
\quad \quad CNM & & & * & * & * \\
\quad \quad Leiden & & & & & \\
\quad \quad Louvain & & & & & * \\
\quad \quad Paris & & * & * & * & * \\
\quad \quad RB & & & & & * \\
\hline 2. Non-modularity based optimization algorithms& 0.01 & 0.1 & 0.3 & 0.5 & 0.7 \\
\quad \quad CPM & * & * & * & * & * \\
\quad \quad GemSec & * & * & * & * & * \\
\quad \quad Genetic Algorithm & * & * & * & * & \\
\quad \quad RE & & & & & \\
\quad \quad Significant scales & * & * & * & * & \\
\quad \quad Surprise & * & * & & & \\
\quad \quad WCC & * & * & * & * & \\
\hline 3. Inferential algorithms & 0.01 & 0.1 & 0.3 & 0.5 & 0.7 \\
\quad \quad BPP & & * & * & * & * \\
\quad \quad BPP with MCMC & & & * & * & * \\
\quad \quad SBM & & * & * & * & * \\
\quad \quad SBM with MCMC & & * & * & * & * \\
\hline 4. Walk-based algorithms & 0.01 & 0.1 & 0.3 & 0.5 & 0.7 \\
\quad \quad Diffusion Entropy Reducer (DER) & * & * & * & * & * \\
\quad \quad Infomap & & & * & * & * \\
\quad \quad Walktrap & & & & & \\
\hline 5. Propagation-based algorithms & 0.01 & 0.1 & 0.3 & 0.5 & 0.7 \\
\quad \quad Chinese whispers & & & * & * & * \\
\quad \quad Asynchronous Label Propagation & * & * & * & * & * \\
\quad \quad Semi-synchronous Label Propagation & * & * & * & * & * \\
\hline 6. Graph cutting algorithms & 0.01 & 0.1 & 0.3 & 0.5 & 0.7 \\
\quad \quad kcut & * & * & * & * & * \\
\quad \quad Keringhan Lin & * & * & * & * & * \\
\hline 7. Markov stability algorithms & 0.01 & 0.1 & 0.3 & 0.5 & 0.7 \\
\quad \quad Markov stability with min NVI & & & & & \\
\quad \quad Markov stability with random & * & * & * & & \\ \hline
\end{tabular}
\label{tab:abcd_ami}
\end{table}

The descriptive rankings of the algorithms in Fig.~\ref{fig:Ranking_AMI} are validated through a statistical analysis in Table~\ref{tab:abcd_ami}. The validation process involved employing a Friedman non-parametric test \citep{friedman1937} to compare multiple groups. Subsequently, a post-hoc Li test for multiple hypotheses \citep{li2008multiple} is utilized. In the statistical evaluation of the AMI results, the null hypothesis of the Friedman test is that the means of the AMIs of two or more algorithms are the same. The Li test compares the control method (PO) with each of the other algorithms. It tests the hypothesis that the mean of the AMIs from the control method is same as that of the other algorithm. Given the multiple testing setting, the Li method controls the family-wise error rate of the comparisons within $0.05$ using a two-step rejection procedure \citep{li2008multiple}.

The validation results for the five experiment settings ($\xi \in\{0.01,0.1,0.3,0.5,0.7\}$) of the ABCD benchmarks are provided in Table~\ref{tab:abcd_ami}. For ABCD graphs with a mixing parameter ($\xi$) set at 0.01, the null hypotheses were rejected in 12 out of 29 post-hoc Li tests. This outcome suggests that, in this experiment setting of $\xi=0.01$, PO's partitions exhibit a statistically significant higher AMI (with the planted partition) when compared to the results of 12 other algorithms. In the case of $\xi=0.1$, the null hypotheses were rejected in 16 out of 29 post-hoc Li tests. While PO does not outperform other algorithms in all comparisons by statistically meaningful margin, it does so for most of\footnote{See the $85$ test rejection asterisks shown for $29 \times 5$ pairwise comparisons in Table~\ref{tab:abcd_ami}.} the comparisons. The difference between the descriptive results in Fig.~\ref{fig:Ranking_AMI} and the statistical results in Table~\ref{tab:abcd_ami} is partly attributed to the limited statistical power in multiple hypothesis testing of 30 algorithms such that the family-wise error rate does not exceed $0.05$ \citep{li2008multiple}.

\subsection{ECS results for retrieving the planted partitions of synthetic benchmarks}
\label{ss:abcd_ECS}

\begin{figure}[htpb!]
 \includegraphics[width=\textwidth]{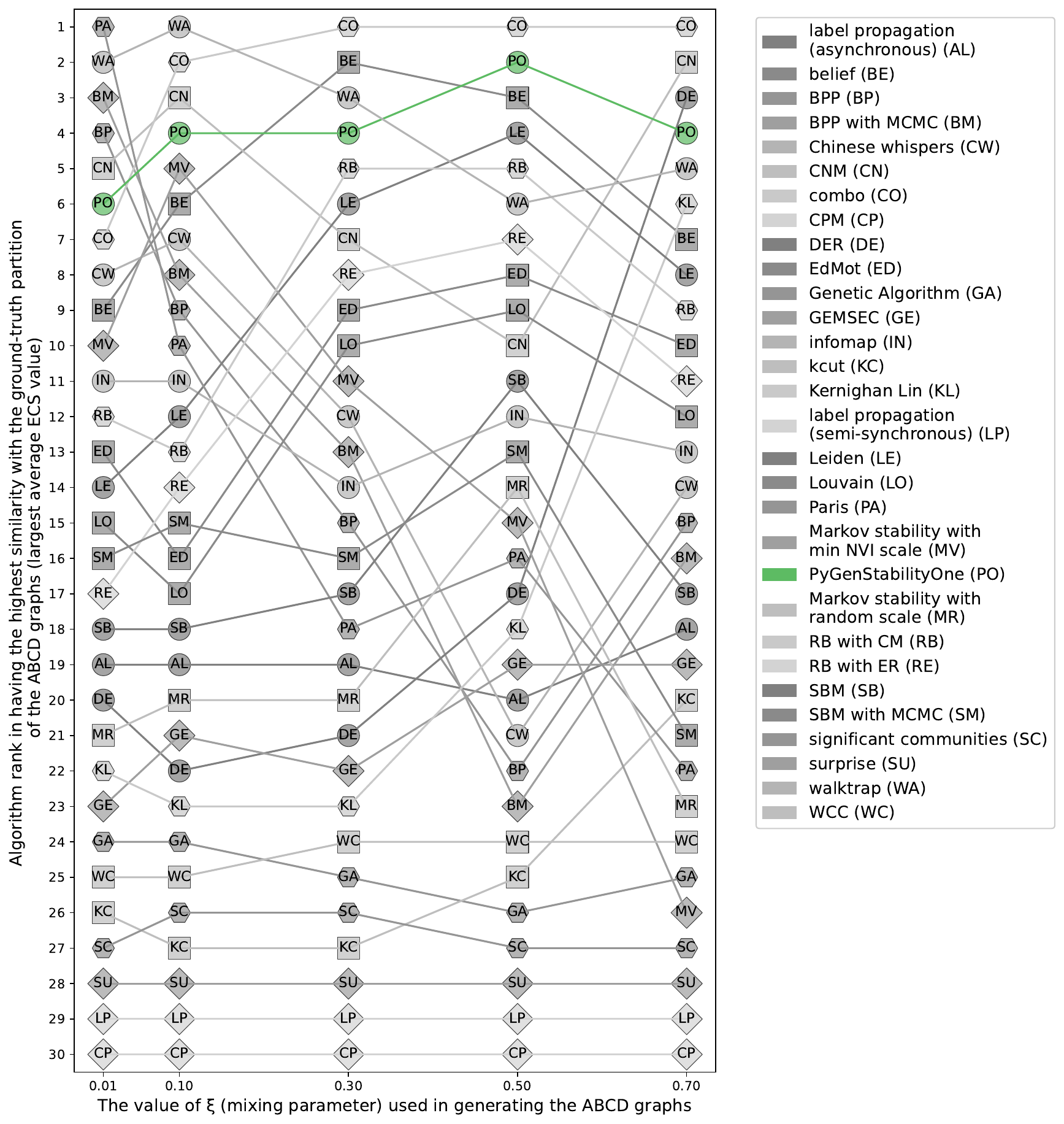}
 \caption{Ranks of 30 CD algorithms based on their average ECS for 100 ABCD graphs in five experiment settings. (Magnify the high-resolution figure on screen for details.)}
 \label{fig:Ranking_ECS}
        \Description[<short description>]{<long description>}
\end{figure}

To ensure that our results are not mere artifacts of using the AMI, Fig.~\ref{fig:Ranking_ECS} shows the ranking of the same 30 algorithms based on the average ECS values (with the planted partitions) for each experiment setting. According to Fig.~\ref{fig:Ranking_ECS}, PO remains in the top six ranks across all experiment settings. There are two other algorithms which have relatively high ECS ranks across the five experiment settings: Combo and Walktrap. Their ranks are always within the top seven (including several settings where they outperform PO). Note that RB and Leiden which had high AMI ranks in Fig.~\ref{fig:Ranking_AMI} show lower performance ranks in Fig.~\ref{fig:Ranking_ECS} where ECS is used as the partition similarity measure.

Similar to the case of AMI ranks, the two algorithms Paris and BM have remarkable ECS ranks when $\xi=0.01$, but their ranks substantially plummet for higher values of the mixing parameter $\xi$. ECS averages of the algorithms are provided in Table~\ref{tab:abcd-avg-ecs} in the appendix.

In Figs.~\ref{fig:Ranking_AMI}-\ref{fig:Ranking_ECS}, both ECS and AMI ranks of the PO algorithm improve generally when the mixing parameter increases. This upward trend comparative performance of PO indicates that it has a relative strength for retrieving planted partitions in the high-noise regime (large mixing parameter $\xi$). For the low-noise regime, however, the multi-scale nature of PO may lead to a partition at a scale that is too fine or too coarse compared to the planted partition. This weakness of PO is manifested in its relatively lower ranks for the low-noise regimes in Figs.~\ref{fig:Ranking_AMI}-\ref{fig:Ranking_AMI} (small mixing parameter $\xi$). For the low-noise regime, the Walktrap algorithm seems to have a particularly strong performance as shown based on AMI and ECS respectively in  Fig.~\ref{fig:Ranking_AMI} and Fig.~\ref{fig:Ranking_ECS}.

The descriptive ECS rankings of the algorithms in Fig.~\ref{fig:Ranking_ECS} can be validated through the same statistical procedure involving the Friedman test \citep{friedman1937} followed by a post-hoc Li test for multiple hypotheses \citep{li2008multiple} with PO as the control method. The validation results for the five experiment settings are provided in Table~\ref{tab:abcd_ecs}. For ABCD graphs with a mixing parameter ($\xi$) set at 0.01, the null hypotheses were rejected in 13 out of 29 post-hoc Li tests. This outcome suggests that, for benchmarks with $\xi=0.01$, PO's partition results exhibit a statistically significant higher ECS when compared to the partitions from 13 other algorithms. In the case of $\xi=0.1$, the null hypotheses were rejected in 12 out of 29 post-hoc Li tests. The results in Table~\ref{tab:abcd_ecs} mean that while PO does not outperform other algorithms by statistically meaningful margins in all comparisons, but it does so for most of the comparisons.\footnote{See the $73$ test rejection asterisks shown for $29 \times 5$ pairwise comparisons in Table~\ref{tab:abcd_ecs}.}

\begin{table}[htbp!]
\centering
\caption{Statistically significant discoveries for comparing ECS values between PO and another algorithm for each value of $\xi$. * in row $a$ and column $\xi$ means that PO has a significantly higher mean ECS compared to the algorithm in row $a$ based on 100 ABCD graphs generated with the mixing parameter $\xi$. }
\begin{tabular}{lccccc}
\hline Algorithm compared to PyGenStabilityOne & $\xi$ & $\xi$ & $\xi$ & $\xi$ & $\xi$ \\ \hline
1. Modularity-based optimization algorithms & 0.01 & 0.1 & 0.3 & 0.5 & 0.7 \\
\quad \quad Belief & & & & & \\
\quad \quad Combo & & & * & & \\
\quad \quad EdMot & & & & & \\
\quad \quad CNM & & & & & \\
\quad \quad Leiden & & & & & \\
\quad \quad Louvain & & & & & \\
\quad \quad Paris & & & * & & * \\
\quad \quad RB & & & & & \\ \hline
2. Non-modularity based optimization algorithms & 0.01 & 0.1 & 0.3 & 0.5 & 0.7 \\
\quad \quad CPM & * & * & * & * & * \\
\quad \quad GemSec & * & * & * & * & * \\
\quad \quad Genetic Algorithm & * & * & * & * & * \\
\quad \quad RE & & & & & \\
\quad \quad Significant scales & * & * & * & * & * \\
\quad \quad Surprise & * & * & * & * & * \\
\quad \quad WCC & * & * & * & * & * \\ \hline
3. Inferential algorithms & 0.01 & 0.1 & 0.3 & 0.5 & 0.7 \\
\quad \quad BPP & & & & * & * \\
\quad \quad BPP with MCMC & & & & * & * \\
\quad \quad SBM & & & * & & \\
\quad \quad SBM with MCMC & * & & * & & * \\ \hline
4. Walk-based algorithms & 0.01 & 0.1 & 0.3 & 0.5 & 0.7 \\
\quad \quad Diffusion Entropy Reducer (DER) & * & * & * & & \\
\quad \quad Infomap & & & & * & \\
\quad \quad Walktrap & & & & & \\ \hline
5. Propagation-based algorithms & 0.01 & 0.1 & 0.3 & 0.5 & 0.7 \\
\quad \quad Chinese whispers & & & & * & * \\
\quad \quad Asynchronous Label Propagation & * & * & * & * & * \\
\quad \quad Semi-synchronous Label Propagation & * & * & * & * & * \\ \hline
6. Graph cutting algorithms & 0.01 & 0.1 & 0.3 & 0.5 & 0.7 \\
\quad \quad kcut & * & * & * & * & * \\
\quad \quad Keringhan Lin & * & * & * & * & \\ \hline
7. Markov stability algorithms & 0.01 & 0.1 & 0.3 & 0.5 & 0.7 \\
\quad \quad Markov stability with min NVI & & & & * & * \\
\quad \quad Markov stability with random & * & * & * & * & * \\ \hline
\end{tabular}
\label{tab:abcd_ecs}
\end{table}

\subsection{Summary of descriptive and statistical results on synthetic benchmarks}
\label{ss:abcd_summary}

In summary, the statistical results, as presented in Tables~\ref{tab:abcd_ami}--\ref{tab:abcd_ecs}, indicate that PO consistently exhibits significantly higher AMI and ECS scores in comparison to the majority of the other 29 algorithms considered. This reinforces the findings from the descriptive rankings shown in Figs.~\ref{fig:Ranking_AMI}--\ref{fig:Ranking_ECS}. In the case of four algorithms EdMot, Leiden, RE, and Walktrap, there exists inadequate evidence to indicate a difference in AMI means or ECS means when compared to PO across all five experimental settings. Walktrap and Combo have high performance in retrieving the planted communities for these ABCD benchmark networks. They are two other recommendable algorithms that stand out in our analysis. PO outperforms 25 out of 29 algorithms considered by statistically meaningful margins based on at least one partition similarity measure (in most cases based on both partition similarity measures). 

Tables~\ref{tab:abcd_ami}--\ref{tab:abcd_ecs} allow us to look at the families of existing algorithms where fewer discoveries (statistical test rejections) were made. Based on AMI values, algorithm families that have fewer test rejection asterisks in Table~\ref{tab:abcd_ami} are modularity-based algorithms and walk-based algorithms. Based on ECS values in Table~\ref{tab:abcd_ecs}, there are fewer discoveries for the same two families of algorithms. This pattern that remains consistent regardless of using AMI or ECS seems to suggest that modularity-based and walk-based methods in our experiments have been generally more successful at retrieving planted partitions compared to the other four families of methods (non-modularity based optimization, inferential, propagation based, and graph cutting).

The descriptive and statistical test results of comparing PO with MR and MV allow us to see the practical relevance of using an ML-based scale selection that we have built upon the existing functionalities of the PyGenStability library. PO outperforms the two naive adaptations of Markov stability (MV and MR) by statistically meaningful margins (at least in one partition similarity measure). This suggests the practical relevance of using our pre-trained scale selection method alongside PyGenStability instead of randomly choosing between PyGenStability's partitions (MR) or selecting the partition that has the lowest normalized variance of information (MV).

\FloatBarrier

\subsection{Assessment of PO's partition scale based on the structure and node labels of real networks}
\label{ss:real}

In this subsection, we illustrate the practical differences that PO makes by comparing its results to the same 29 baseline algorithms on five real networks from different context and domains. We evaluate the similarity between the partitions generated by each of the 30 algorithms and node labels (discrete-valued attributes) across three real benchmark (risk, dolphins, and football) networks. We also use two other real networks (wiki and netsci) because they have a context that is suitable for discussing the scale of partitions. These two real network do not have node labels.

Table~\ref{tab:real} provides the names and references for each of these five networks as well as their sizes and orders. Before using any of the algorithms, we quantify the extent to which the three networks have well-separated communities induced by node labels (assortative mixing by node labels). For this sanity check, we calculate the modularity $Q$ for each network based on the partition corresponding to the node labels. Table~\ref{tab:real} provides the modularity values for the three networks that have node labels. These three networks are frequently referenced in the literature \citep{hric2014community,yang2015defining,newman2016structure,sobolevsky_optimality_2017,kawamoto2018comparative,chen_global_2018,edmot_2019,sobolevsky2022gnn} for this purpose because their class labels form reasonably well-separated communities.

\begin{table}[ht]
 \centering
 \caption{Real networks used for discussing the results of the PO algorithm}
 \begin{tabular}{cccc} \hline 
 Network dataset and reference& $n$& $m$ & $Q_{\text{(node label partition)}}$\\ \hline 
 Risk game (risk) \citep{steinhaeuser2010identifying}& 41 & 83 & 0.62 \\ 
 Dolphins \citep{lusseau2003bottlenose}& 62 &	159 & 0.52\\ 
 American college football (football) \citep{girvan2002community}& 115 &	613 & 0.55\\ 
 Wikipedia fields of science (wiki) \citep{Calderone2020} & 687 & 6523 & NA\\ 
 Network science co-authorship (netsci) \citep{newman_finding_2006} & 1589	& 2742 & NA \\ \hline
 \end{tabular}
 \label{tab:real}
\end{table}

CD algorithms are not designed to retrieve node labels \citep{peel2017ground}, and network structure may not strongly correlate with arbitrarily selected node attributes \citep{farber_using_2010}. Thus, we refrain from regarding a specific set of node metadata as ground truth. Nevertheless, comparing partitions yielded by various algorithms against node metadata \citep{yang2015defining,newman2016structure} proves useful for assessing them in a real-world context without synthetic benchmarks. We discuss the results on the five networks of Table~\ref{tab:real} in the five Sections~\ref{sss:risk}--\ref{sss:netsci}.

\begin{figure}[!htbp]
 \centering
 \begin{subfigure}[b]{0.4\textwidth}
 \centering
 \includegraphics[width=\textwidth]{figures/risk_pygenone.pdf}
 \caption{Risk network, $c=7$ \citep{steinhaeuser2010identifying}}
 \label{subfig:risk}
 \end{subfigure}
 \centering
 \begin{subfigure}[b]{0.4\textwidth}
 \centering
 \includegraphics[width=\textwidth]{figures/dolphins_pygenone.pdf}
 \caption{Dolphins network, $c=5$ \citep{lusseau2003bottlenose}}
 \label{subfig:dolphins}
 \end{subfigure}
 \begin{subfigure}[b]{0.49\textwidth}
 \centering
 \includegraphics[width=\textwidth]{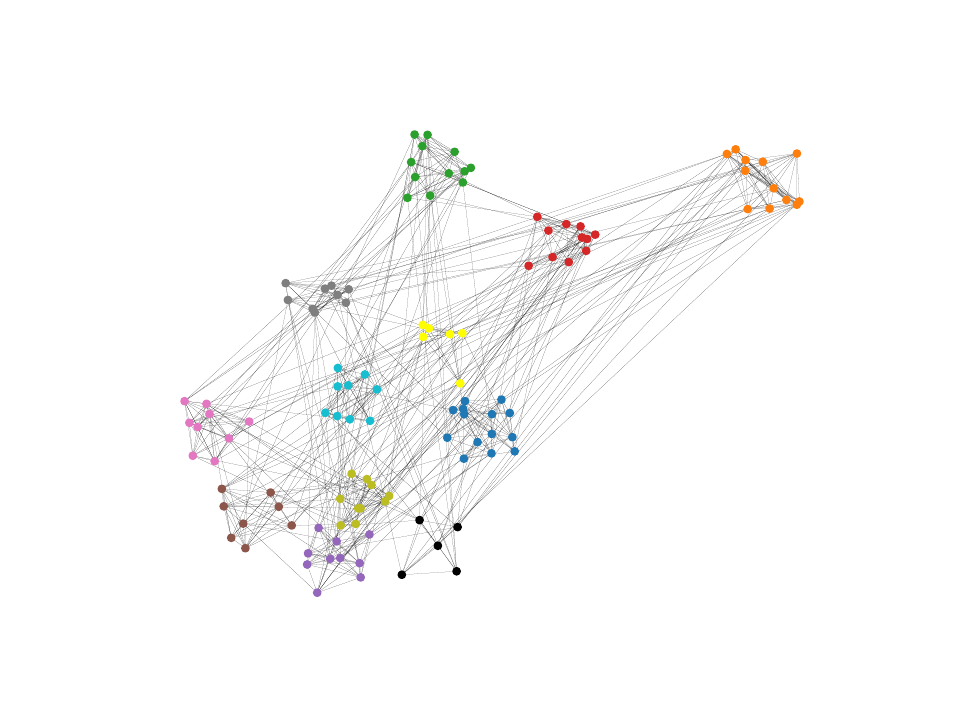}
 \caption{Football network, $c=12$ \citep{girvan2002community}}
 \label{subfig:football}
 \end{subfigure}
 \begin{subfigure}[b]{0.49\textwidth}
 \centering
 \includegraphics[width=\textwidth]{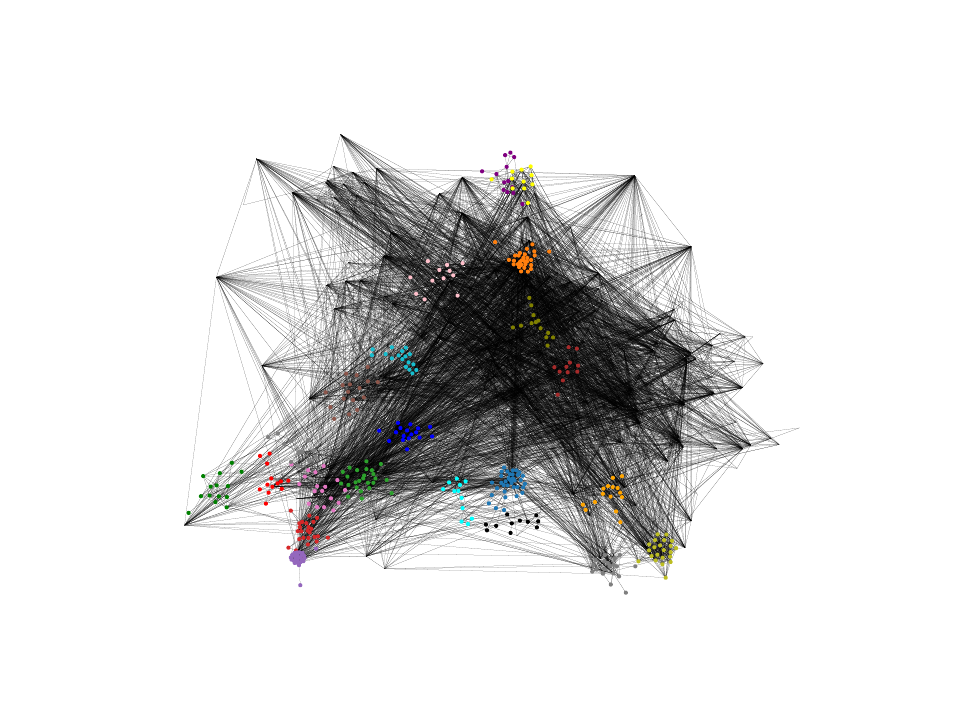}
 \caption{Wiki Science\citep{Calderone2020}, $c=20$ communities and a considerable number of outlier nodes}
 \label{subfig:wikiscience}
 \end{subfigure}
 \begin{subfigure}[b]{0.48\textwidth}
 \centering
 \includegraphics[width=\textwidth]{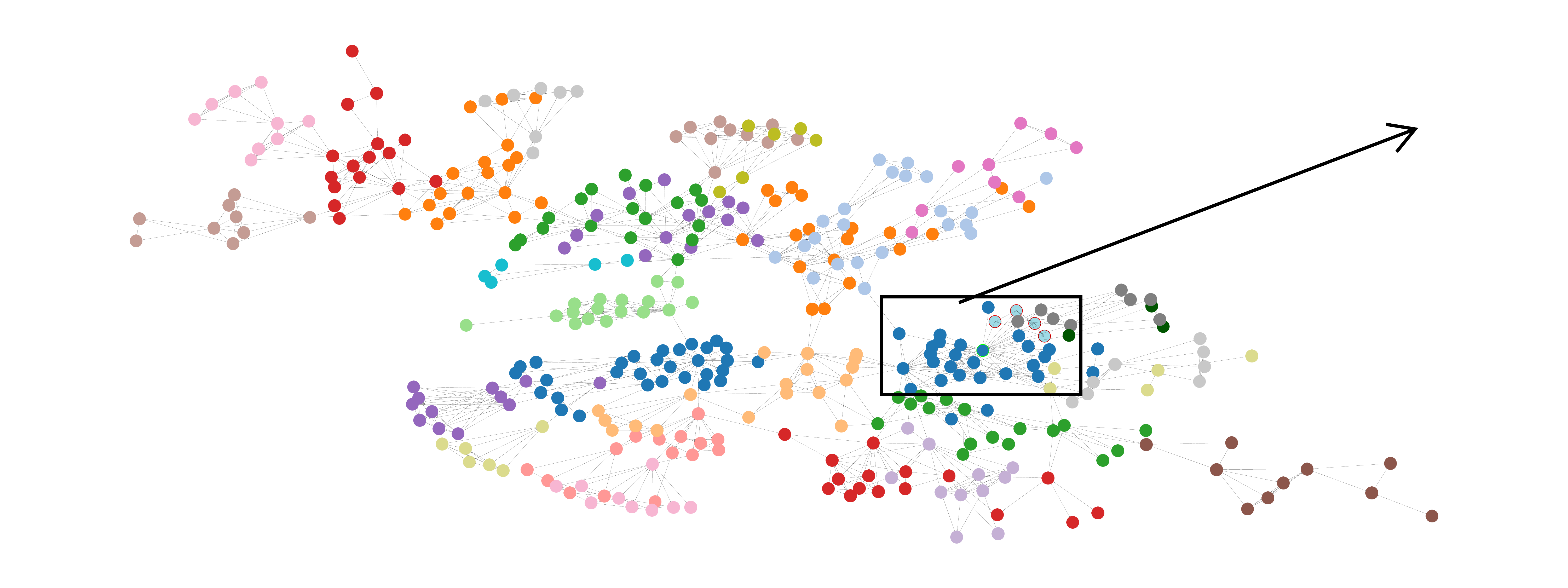}
 \caption{Network science co-authorship network, $c=30$ \citep{newman_finding_2006}}
 \label{subfig:authorship}
 \end{subfigure}
 \begin{subfigure}[b]{0.48\textwidth}
 \centering
 \includegraphics[width=\textwidth]{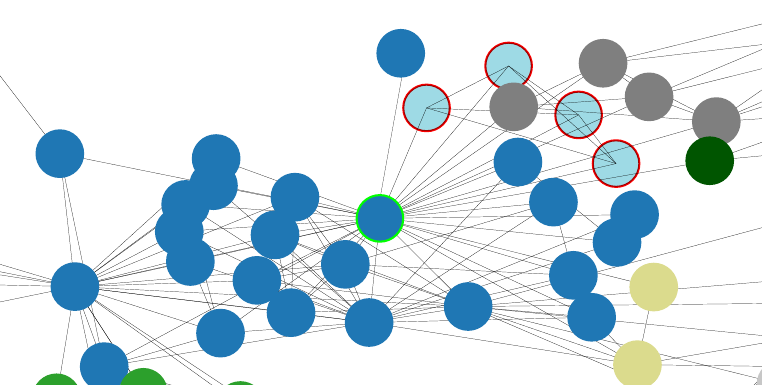}
 \caption{A magnified of the Network science co-authorship network with five nodes shown with distinctive outline colors.}
 \label{subfig:authorship_zoom}
 \end{subfigure}
 \caption{Partitions obtained by PyGenStabilityOne (PO) on five networks (a)-(e). Node colors represent the communities according to the PO's partition. $c$ indicates the number of communities in the partition. Magnify the high-resolution color figure on screen for more details. }
 \label{fig:real}
        \Description[<short description>]{<long description>}
\end{figure}

\subsubsection{Risk game}
\label{sss:risk}
The first network is an illustrative example for community detection which is created from the territories in a board game called \textit{Risk} \citep{steinhaeuser2010identifying}. In this small network, nodes represent territories, and the edges represent the connections between the territories in the Risk board game. The partition obtained by PO on this network is made up of seven communities shown by colors in Fig.~\ref{subfig:risk}. This network has node labels which represent the continents of the territories. The partition produced using PO is highly similar to the node labels as indicated by both partition similarity values AMI=0.928012 and ECS=0.857143. Except for the Infomap algorithm which has partitions similarity values of AMI=0.944597 and ECS=0.935941 with the node label partition, all other algorithms return partitions that have lower similarity with the node label partition, in comparison with PO. 

\subsubsection{Social network of dolphins}

The second network is a real social network of dolphins. It is based on frequent associations (edges) between 62 bottlenose dolphins (nodes) which were collected in a behavioral ecology field study in New Zealand \citep{lusseau2003bottlenose}. The partition that PO returns for this network has five communities as shown in Fig.~\ref{subfig:dolphins}, and has a modularity of 0.52 (equal to the modularity of the node label partition). This partition is highly similar to the node labels as indicated by the partition similarity measures AMI=0.857713 and ECS=0.830005. The node label partition also has five communities which were assigned by the behavioral ecologists to these dolphins. Except for Combo, the partitions from all other 28 algorithms have lower similarities with the node label partition (their AMI and ECS values are all below the corresponding values for PO). The Combo algorithm returns a partition involving 4 communities which has partition similarity values of AMI=0.867925 (higher than PO's) and ECS=0.828853 (lower than PO's) with the node label partition.

\subsubsection{Network of American college football teams}

The third network is based on data on matches (edges) between American college football teams (nodes) in Division IA colleges during the regular season Fall 2000. PO returns a partition for this network that is made up of 12 communities which is shown in Fig.~\ref{subfig:football}. This network has node labels which indicate the leagues (NCAA conferences) that each team belongs to \citep{girvan2002community}. As teams are more likely to have matches with other teams of the same league, there is an association between the structure and the partition corresponding to the node labels. There are also 12 node labels, and their corresponding partition is highly similar to the partition obtained by PO as indicated by the partition similarity measures AMI=0.899167 and ECS=0.866791. The modularity for the partition of PO is 0.6 showing that the obtained communities are reasonably well-separated (also visible in Fig.~\ref{subfig:football}). None of the other 29 algorithms produces a partition as similar as PO's partition to the node label partition of this network. 

\subsubsection{Network of similarity between field of science on Wikipedia}

Moving on to the networks without node labels, the fourth example is an informational network of the fields of science (nodes) and the cosine similarity between the content of their English Wikipedia pages as in early 2020. This network was created from pairwise similarity of web pages such that only the edges whose similarity exceeded a z-score threshold of 1.96 were included \citep{Calderone2020}. The partition obtained from PO is shown in Fig.~\ref{subfig:wikiscience}. It is made up of 20 communities and a considerable number of outliers which in turn have reduced the modularity to 0.29. While a partition that reflect the groups of node labels does not exist for this network, the communities obtained by PO can be inspected for relevance. Upon close inspection of these communities, fields in the same community are seen to be reasonably relevant to each other. For example, one of the communities is made up of the following fields: 
`Environmental science',
 `Atmospheric physics',
 `Atmospheric science',
 `Climatology',
 `Environmental science',
 `Geomorphology',
 `Geophysics',
 `Hydrogeology',
 `Hydrology',
 `Paleoclimatology',
 `Paleontology',
 `Seismology',
 `Climatology',
 `Geomorphology',
 `Paleoclimatology', and
 `Physical geography'. These are relevant to each other and arguably different from other fields of physics. Note that PO returns these communities without any input from the user w.r.t.\ the scale, the resolution, or the number of communities. The structure of the input graph is summarized by features which automatically select a scale that results in the above-mentioned fields being clustered together, while not lumped into other fields of physics for example.

\subsubsection{Co-authorship network of researchers in the field of network science}
\label{sss:netsci}
The last network that we use for evaluating PO is a co-authorship network representing researchers who have published in the field of network science. This network is the largest connected component of a one-mode projection from the bipartite graph of authors and their scientific publication as of 2006 \citep{newman_finding_2006}. The partition from PO is shown in Fig.~\ref{subfig:authorship} and is made up of 30 communities. The modularity for this partition is 0.83 indicating that the communities are well-separated. There is no metadata for this dataset that reflects the grouping of node labels. However, we closely inspect the communities obtained from PO to see if they represent intuitive clusters of authors who have published together while maintaining the desirable division between researchers who have not published together (as of 2006).

To discuss the difference that PO makes, we focus on a specific part of the network which is magnified in Fig.~\ref{subfig:authorship_zoom}.
The community of four nodes shown in light blue with red outline color is made up of the following researchers: `BERLOW, E', `DUNNE, J', `WILLIAMS, R', and `MARTINEZ, N'. Upon searching these names, we obtain a 2002 article entitled ``Two degrees of separation in complex food webs'' \citep{williams2002two} that is co-authored by these researchers. This is as if PO has reverse engineered the data generation process; because it inductively generates a cluster that represents a separate entity from which these few pieces of network data (the four nodes and their six edges to each other) are generated. Note that the same publication is also co-authored by `BARABASI, A' (shown by a dark blue node and light green outline color in the centre of Fig.~\ref{subfig:authorship_zoom}) who is a prolific researcher in the field of network science. Despite the co-authorship ties to `BARABASI, A', the PO algorithm has not lumped the four researchers `BERLOW, E', `DUNNE, J', `WILLIAMS, R', and `MARTINEZ, N' into the community of `BARABASI, A' and other network scientists (the community of dark blue nodes). This result is aligned with the additional information that is publicly available about these five researchers: `BERLOW, E', `DUNNE, J', `WILLIAMS, R', and `MARTINEZ, N' are all ecologists (according to their public profiles and the topics of their publications as of 2006). They have not co-authored with other co-authors of `BARABASI, A' as of 2006 (except for each other). Therefore, it is arguably consistent with the co-authorship patterns (the local structure of the network depicted in Fig.~\ref{subfig:authorship_zoom}) that the four ecologists are assigned into a community of four that is separate from the large dark blue community of network science researchers including `BARABASI, A'.

We run the other 29 CD algorithms on this network, and observe that the partitions returned are considerably different from the partition of PO. In the partitions obtained by most other algorithms (all algorithms except for CW, SC, GA, WC, and SU), the four above-mentioned ecologist researchers are lumped together with the large dark blue community of network scientists including `BARABASI, A'. The five algorithms CW, SC, GA, WC, and SU, separate the four ecologists from the network scientists. However, this comes at the cost of increasing the number of communities to 63, 96, 120, 139, and 378 respectively. Using PO leads to the desirable separation while the number of communities is 30 (and modularity is 0.83), which seems more reasonable for the structure of the network in Fig.~\ref{subfig:authorship}.

One may argue that a practitioner who has exogenous information about a network can arbitrarily tweak the user parameters of some of these existing CD algorithms and obtain communities that are consistent with their intuition. This is a commonplace practice for better or worse (we discuss this point further in Section \ref{ss:cscw} by reviewing some examples from the literature). Unlike other algorithms, PO returns one robust partition according to the scale that has an explainable association with the features that represent the structure of the input network. The advantage of PO is not only in the structurally intuitive partitions it produces as seen in this section. It is in the automatic process of using the structure of the input graph to inform the multi-scale community detection task which obviates the need from any arbitrary tinkering.

\subsection{Empirical scalability comparison for the 30 CD algorithms}
\label{ss:time}

So far, we compared PyGenStabilityOne (PO) and the other 29 algorithms based on their retrieval of planted partitions and their performance on real networks. In this section, we use the same 500 ABCD graphs generated in Section \ref{ss:generate_retrieval} to compare the scalability of the 30 algorithms. We measure the empirical run time for each of the 30 algorithms on all the 500 networks. 

\begin{figure}
 \centering
\includegraphics[width=0.975\textwidth]{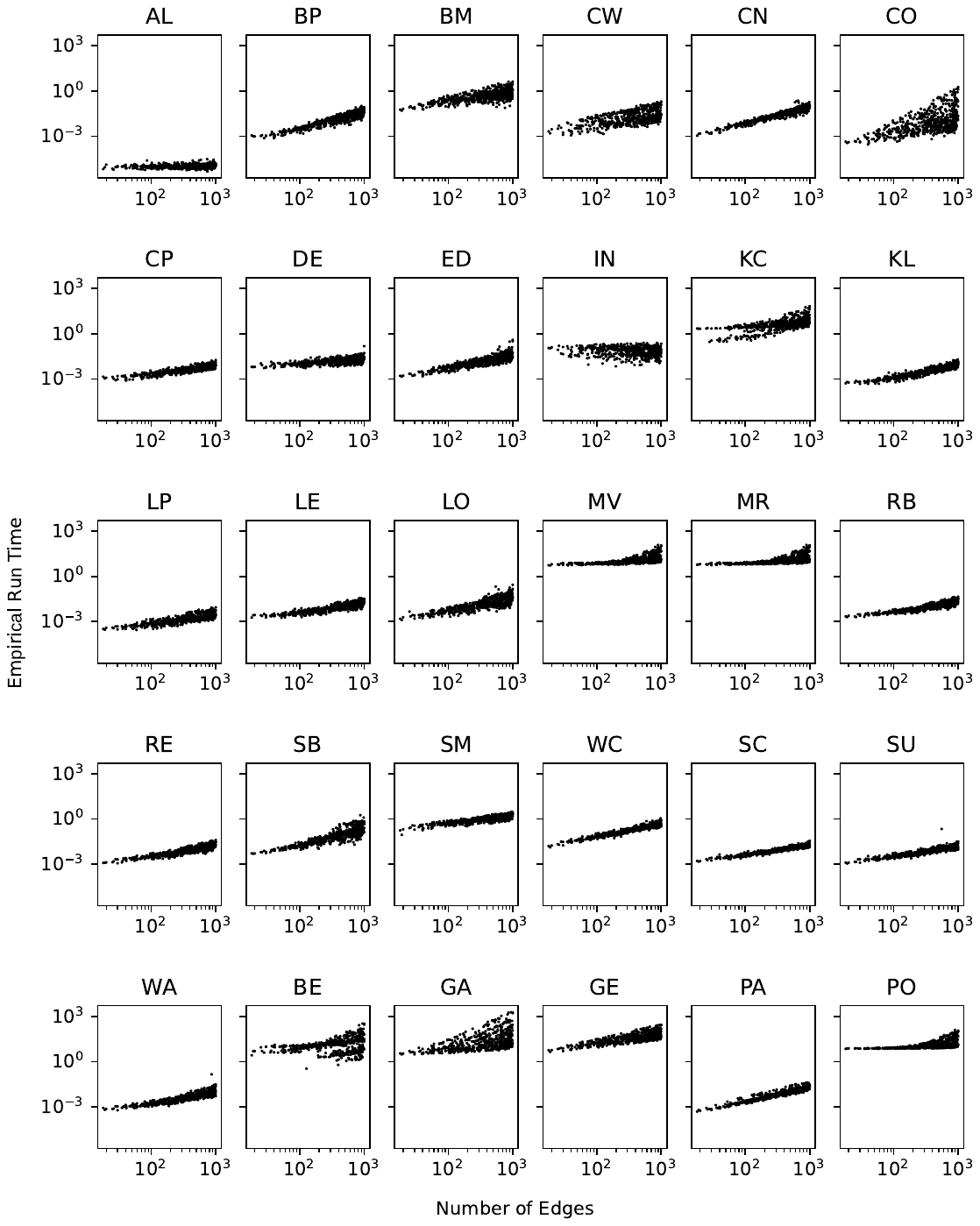}
 \caption{Run time of each of the 30 algorithms (y-axis) and the instance graph size (x-axis) shown in a panel as a log-log scatterplot.}
 \label{fig:scalability}
        \Description[<short description>]{<long description>}
\end{figure}

Fig.~\ref{fig:scalability} shows a log-log scatterplot of empirical solve time and instance graph size for each of the 30 algorithms. It can be observed in Fig.~\ref{fig:scalability} that there is a wide range of variation between how the run times scale with instance size. The label propagation algorithms (both versions), CPM, Leiden (and its derivatives RB and RE), significant scales, and Paris show the best scalability among the 30 algorithms. The three slowest algorithms which also do not scale well when the instance size increases are Belief, genetic algorithm, and GemSec. PO lies somewhere in the middle w.r.t.\ the empirical run time mostly taking a few seconds per instance for each network within this range of instance sizes. Not that we have included all eight steps of the PO algorithm (described in Section~\ref{s:proposed}) in its run time measurement. Therefore, it is not expected for PO or an algorithm of comparable sophistication to ever outperform CD algorithms which were designed mainly for scalability to large-scale instances (e.g. label propagation). Despite its longer time per run, a distinctive silver lining of PO's practical efficiency is that it is sufficient to run it once and once only for obtaining one definite robust partition at a suitable scale, rather than spending actual time tweaking its output through multiple runs.

Taking the comparative results from Sections \ref{fig:Ranking_AMI}--\ref{fig:Ranking_ECS} and the comparative running times in Fig.~\ref{fig:scalability} together, we arrive at a precious few algorithms which offer practical run time and reliable retrieval rates. These algorithms include PO, Leiden, Combo, RB, RE, and Walktrap. Using scalability as the sole criterion for choosing an algorithm leads to choosing a constant-time algorithm that gives a random partition for any input. So, we argue that scalability can only be assessed alongside the functionality of the CD algorithm.
There are algorithms, like asynchronous label propagation, that are substantially faster and more scalable than the six above-mentioned algorithms. However, our combined results in Figs.~\ref{fig:Ranking_AMI}--\ref{fig:Ranking_ECS} and Fig.~\ref{fig:scalability} show that speed and scalability of CD algorithms generally come at a cost in lower retrieval rate. Therefore, we argue that the six algorithms PO, Leiden, Combo, RB, RE, and Walktrap which strike a balance between functionality and time efficiency can be more justifiable than the others in standard use cases. For large-scale networks, obtaining a robust partition through multi-resolution community detection (including PO) is more costly than a single run of a fast and scalable method. Ultimately, it is in the discretion of the user to deploy our comparative results for deciding the most accurate and cost-effective method for their use case. Our results demonstrate the three algorithms GA, GemSec, and Belief as the slowest algorithms which also have some of the lowest retrieval rates (regardless of using AMI or ECS). 

\section{Community detection in current and future CSCW research}
\label{ss:cscw}

Due to the relevance of both interdependent network data and unsupervised clustering to some areas of \textit{Computer-Supported Cooperative Work \& Social Computing} (CSCW) such as social media research, community detection has been used in previous CSCW research \citep{cscw14_community,datta2017identifying,cscw21_community}. PO offers a more accurate and stable (see Figs.~\ref{fig:Ranking_AMI}--\ref{fig:Ranking_ECS}) method for applied social network and social computing research. It also offers a more straightforward analysis pipeline which does not require any arbitrary choices or ad-hoc manipulations which may impact the results. To demonstrate this advantage of PO, we compare and contrast it against the existing practices of using CD algorithms exemplified in previous CSCW research \citep{cscw14_community,datta2017identifying,cscw21_community}.

The research in \citep{datta2017identifying} aims to identify misaligned inter-group links in online social networks based on three types of networks \textit{author}, \textit{term}, and \textit{similarity} networks. The community detection choices made in \citep{datta2017identifying} have been reported particularly well. The authors have described their usage of CD algorithms as follows. First, six different CD algorithms (CNM, Infomap, label propagation, Louvain, Walktrap, and the spinglass algorithm) are applied on some of the networks (\textit{author} and \textit{term} networks). Then, Infomap and Walktrap are found to produce too many small communities, while label propagation and CNM are observed to produce one very large community (the practical problem that PO tackles precisely). Louvain was found to produce the desirable number of reasonably-sized communities on those networks. This comparison of sizes (i.e.\ scales) of communities was used to decide on using the Louvain algorithm on other networks (\textit{similarity} networks) in \citep{datta2017identifying}. While the process for choosing the CD algorithm can be viewed as somewhat ad-hoc, the choice of the algorithm has major consequences for the resulting partitions (as we demonstrated in Sections~\ref{ss:visual} and \ref{ss:real}).

Other studies offer interesting CSCW applications for community detection \citep{cscw14_community,cscw21_community}. Bhattacharya et al.\ have used community detection in a study that proposes a method for identifying topical groups in Twitter \citep{cscw14_community}. 
In another study based on Twitter, Bandy and Diakopoulos use community detection in a longer analysis pipeline to study the algorithmic gatekeeping in Twitter \citep{cscw21_community}. They characterize the effects of Twitter's algorithmic timeline curation on source diversity (increase) and topic diversity (mixed effect). The authors describe that they accidentally get the eight communities that align with their other results, and explain that otherwise they might have ``adjusted the modularity parameter in the Louvain algorithm to explore alternative clusterings of communities" \citep[pp 78:7]{cscw21_community}. We do not scrutinize these CSCW studies to cynically single out their honorable authors or put their research choices under criticism. We describe some aspects of their work to exemplify a commonplace and improvable practice in using CD algorithms for which we are proposing the much-needed solution. We appreciate authors of \citep{cscw14_community,datta2017identifying,cscw21_community} for clear and transparent documentation of their methodological choices w.r.t. CD.

Choosing the CD algorithm is sometimes overlooked, especially when the community detection is only one step among several steps of a longer analysis pipeline. Only the Louvain algorithm has been used in these two empirical CSCW research studies \citep{cscw14_community,cscw21_community} without a selection criterion or a reference to any comparison of the existing algorithms. It can be argued that if a selection was made among existing CD algorithms, their CD results could have been different. This highlights that our comparison of 30 CD algorithms can be relevant and useful for future CSCW research so that more accurate and reliable methods can be adopted for empirical research. Our comparison is also useful for future methodological research in CSCW so that the most accurate and reliable algorithms can be chosen as baselines to compare to the proposed method.

In the context of CSCW applications, the key practical features of our proposed algorithm are as follows: (1) PO does not require any assumptions or manipulations from the user such as an untuned hyperparameter or a resolution parameter whose optimal value is left to the practitioner. (2) Unlike heuristic methods whose results depend on the random number seed and are often re-run numerous times in the hope of getting better results, our proposed algorithm only requires one run to return a definite robust partition at a scale aligned with the graph structural properties. (3) Unlike PyGenStability, it returns one partition rather than a sequence of multiple partitions at different scales for the practitioner to choose from. Taken together with its accuracy and stability over different levels of noise, these practical features of PO make it a suitable computational tool to reliably handle community detection.

\section{Discussion}
\label{s:discuss}

The seven comparative analyses provided in Sections~\ref{ss:visual}--\ref{ss:time} underscore the practical advantages of using our proposed algorithm, PyGenStabilityOne (PO), when contrasted with 29 alternative approaches for community detection. 

Our comparison is based on the default hyperparameter settings of 29 baseline algorithms. Some of them do not have any hyperparameters to be adjusted by the user. One may argue that our comparison could have been more informative if algorithms that have hyperparameters are somehow optimized to their non-default hyperparameter settings. We consider our use of default settings as a limitation of our comparison. Despite the limitation, using default settings makes our comparison reproducible and more reflective of the common usage of these algorithms by the average user (who is more likely to use default settings).

PO has a statistically meaningful advantage over 25 other CD algorithms in retrieving the planted communities of benchmark graphs. With the exception of Paris and SC, these reviewed algorithms are based on a single scale (single resolution). They leave much to be desired in taking full advantage of the structural features of the input network. PO works based on a multi-scale Markov stability framework and takes the structural features of the input network into account for its automatic scale detection. These two differences explain why PO outperforms 25 other algorithms.

PO is also capable of returning intuitive partitions on real networks showing high alignment with their node label partitions and the structure of real networks. The partition obtained by PO does not require any arbitrary adjustments or any additional assumptions on the user's part; it is robust and at a single scale that is informed by the structure of the input graph. As the scale is automatically determined by the PO algorithm, the number and size of communities will also become reasonable values as demonstrated in seven visualizations of real networks from different contexts in Sections~\ref{ss:visual} and \ref{ss:real}.

While it may be tempting to interpret the results as suggesting Markov stability to be a superior objective function compared to other objective functions for CD, such an interpretation will not be accurate due to the confounding of other factors that differ between the algorithms. Future research may explore multi-scale approaches for CD algorithms that rely on other objective functions. It is hoped that our detailed presentation of the PO algorithm will facilitate the development of methods that outperform PO.

On the methodological side, our contributions were twofold: (1) demonstrating the advantages of using a simple pre-trained predictive model for multi-scale Markov stability, and (2) reaffirming the practical relevance of robust partitions obtained through multi-scale Markov stability compared to 29 other methods. On the empirical side, PO addresses a practical challenge in social network analysis and social computing, thereby enhancing a widely used computational tool for studying networked systems of social relevance.

\section*{Software availability and license}
The PyGenStabilityOne (PO) algorithm is publicly available at \url{https://github.com/saref/PyGenStabilityOne}. The use terms of PO are based on the GNU General Public License v3.0 \url{https://gnu.org/licenses/gpl-3.0.en.html}. Permissions of this strong copyleft license are conditioned on making available complete source code of licensed works and modifications, which include larger works using a licensed work, under the same license.

\begin{acks}
Authors acknowledge the anonymous reviewers and meta-reviewers of CSCW for helpful comments which have improved the quality of this manuscript over its earlier versions. Authors do not have any conflicts to declare. As the name suggests, PyGenStabilityOne relies on PyGenStability which does most of the heavy lifting behind the results reported in this manuscript. Authors acknowledge the team of PyGenStability (Alexis Arnaudon, Dominik J. Schindler, Robert L. Peach, Adam Gosztolai, Maxwell Hodges, Michael T. Schaub, and Mauricio Barahona) for making their software publicly available. This study has been supported by the Data Sciences Institute at the University of Toronto.
\end{acks}



\appendix

\section{Additional numerical results}

\subsection{Average AMI for each of the 30 CD methods in each of the five ABCD experiment settings}

\begin{table}[ht!]
\small
\centering
\caption{Average AMI for each of the 30 CD algorithms in each of the five ABCD experiment settings (each value of $\xi$)}
\begin{tabular}{llllll}
\hline
Algorithm \textbackslash \ $\xi$ & 0.01 & 0.1 & 0.3 & 0.5 & 0.7 \\ \hline
PyGenStabilityOne & 0.920493 & 0.888722 & 0.731668 & 0.492260 & 0.178296 \\
belief & 0.908730 & 0.872955 & 0.667745 & 0.402757 & 0.045899 \\
BPP & 0.907077 & 0.781043 & 0.473382 & 0.043888 & 0.000000 \\
BPP with MCMC & 0.933735 & 0.806711 & 0.507249 & 0.043537 & 0.000007 \\
Chinese whispers & 0.917539 & 0.887096 & 0.645010 & 0.098250 & 0.020156 \\
Combo & 0.919884 & 0.906193 & 0.763176 & 0.497164 & 0.170715 \\
CPM & 0.000000 & 0.000000 & 0.000000 & 0.000000 & 0.000000 \\
DER & 0.619070 & 0.524286 & 0.398007 & 0.215354 & 0.078891 \\
EdMot & 0.863978 & 0.845659 & 0.684886 & 0.450476 & 0.163103 \\
Genetic algorithm & 0.706952 & 0.670259 & 0.416827 & 0.217894 & 0.139202 \\
GemSec & 0.653493 & 0.638763 & 0.482151 & 0.269096 & 0.107797 \\
CNM & 0.927765 & 0.878954 & 0.675435 & 0.372664 & 0.135652 \\
infomap & 0.866716 & 0.830611 & 0.670542 & 0.378866 & 0.102234 \\
kcut & 0.000815 & -0.000386 & 0.000036 & 0.001158 & -0.000517 \\
Keringhan Lin & 0.522966 & 0.460165 & 0.375014 & 0.226921 & 0.073444 \\
Asynchronous label propagation & 0.787320 & 0.747934 & 0.612395 & 0.235895 & 0.069052 \\
Semi-synchronous label propagation & 0.823490 & 0.764068 & 0.573115 & 0.167497 & 0.051840 \\
Leiden & 0.921773 & 0.901337 & 0.734870 & 0.470164 & 0.163958 \\
Louvain & 0.921400 & 0.877696 & 0.713213 & 0.453911 & 0.147747 \\
Markov stability with min NVI & 0.907128 & 0.871891 & 0.705714 & 0.412507 & 0.153342 \\
Markov stability with random & 0.716248 & 0.745573 & 0.604303 & 0.421414 & 0.162807 \\
Paris & 0.964963 & 0.799557 & 0.511702 & 0.265896 & 0.108647 \\
RB & 0.922838 & 0.901026 & 0.747405 & 0.498813 & 0.156621 \\
RE & 0.909186 & 0.899525 & 0.736806 & 0.481951 & 0.151930 \\
SBM & 0.844630 & 0.709565 & 0.456509 & 0.069812 & -0.006576 \\
SBM with MCMC & 0.850792 & 0.702418 & 0.440768 & 0.044859 & -0.005510 \\
significant scales & 0.586357 & 0.595749 & 0.494496 & 0.329501 & 0.156609 \\
surprise & 0.793681 & 0.788261 & 0.640432 & 0.430844 & 0.182222 \\
Walktrap & 0.920032 & 0.899926 & 0.745977 & 0.475442 & 0.170127 \\
WCC & 0.520103 & 0.524841 & 0.474244 & 0.346609 & 0.159600 \\ \hline
\label{tab:abcdavg}
\end{tabular}
\end{table}

\clearpage

\subsection{Average ECS for each of the 30 CD algorithms in each of the five ABCD experiment settings}

\begin{table}[ht!]
\small
\centering
\caption{Average ECS for each of the 30 CD algorithms in each of the five ABCD experiment settings (each value of $\xi$)}
\begin{tabular}{llllll}
\hline
Algorithm \textbackslash \ $\xi$ & 0.01 & 0.1 & 0.3 & 0.5 & 0.7 \\ \hline
PyGenStabilityOne & 0.88244 & 0.830603 & 0.683084 & 0.483353 & 0.250882 \\
belief & 0.861263 & 0.823718 & 0.692505 & 0.478881 & 0.238575 \\
BPP & 0.927019 & 0.796998 & 0.588966 & 0.254483 & 0.21867 \\
BPP with MCMC & 0.944736 & 0.799716 & 0.594337 & 0.254252 & 0.218665 \\
Chinese whispers & 0.868784 & 0.820031 & 0.607106 & 0.259253 & 0.22167 \\
Combo & 0.881954 & 0.863775 & 0.72142 & 0.51182 & 0.272727 \\
CPM & 0.03297 & 0.045773 & 0.054183 & 0.068147 & 0.09055 \\
DER & 0.556774 & 0.432611 & 0.389808 & 0.303491 & 0.253164 \\
EdMot & 0.793994 & 0.773446 & 0.623698 & 0.425078 & 0.236053 \\
Genetic algorithm & 0.3939 & 0.337007 & 0.253032 & 0.176491 & 0.181347 \\
GemSec & 0.46182 & 0.475661 & 0.384416 & 0.275599 & 0.211638 \\
CNM & 0.89889 & 0.839676 & 0.633448 & 0.382745 & 0.255452 \\
infomap & 0.820763 & 0.787997 & 0.590105 & 0.363299 & 0.227488 \\
kcut & 0.290443 & 0.223798 & 0.228685 & 0.210915 & 0.209744 \\
Keringhan Lin & 0.488825 & 0.363112 & 0.347486 & 0.279629 & 0.240825 \\
Asynchronous label propagation & 0.689402 & 0.620971 & 0.519654 & 0.267025 & 0.213446 \\
Semi-synchronous label propagation & 0.03297 & 0.045773 & 0.054183 & 0.068147 & 0.09055 \\
Leiden & 0.792873 & 0.783796 & 0.647882 & 0.461558 & 0.237263 \\
Louvain & 0.792009 & 0.771023 & 0.618816 & 0.41201 & 0.228992 \\
Markov stability with min NVI & 0.83661 & 0.82789 & 0.615754 & 0.327179 & 0.166282 \\
Markov stability with random & 0.548784 & 0.596701 & 0.474689 & 0.34037 & 0.195411 \\
Paris & 0.968609 & 0.790849 & 0.525602 & 0.306537 & 0.207948 \\
RB & 0.79475 & 0.781463 & 0.649333 & 0.453439 & 0.236253 \\
RE & 0.766179 & 0.77727 & 0.624397 & 0.443089 & 0.23426 \\
SBM & 0.752188 & 0.741012 & 0.540179 & 0.379549 & 0.216595 \\
SBM with MCMC & 0.787378 & 0.774997 & 0.562135 & 0.361674 & 0.208722 \\
significant scales & 0.272353 & 0.250485 & 0.230936 & 0.168902 & 0.155801 \\
surprise & 0.033926 & 0.046789 & 0.055307 & 0.069425 & 0.092532 \\
Walktrap & 0.953414 & 0.887239 & 0.690934 & 0.447648 & 0.247238 \\
WCC & 0.309614 & 0.296879 & 0.30096 & 0.250082 & 0.189879 \\ \hline
\label{tab:abcd-avg-ecs}
\end{tabular}
\end{table}

\end{document}